\documentstyle[12pt,aj_pt4]{article}

\slugcomment{Submitted to AJ}

\lefthead{Fulbright \& Kraft}
\righthead{Oxygen Abundances}

\begin{document}{
\title{Oxygen Abundances in Two Metal-poor Subgiants\\ from the Analysis of the
$\lambda$ 6300 \AA$\;$ Forbidden O I Line}

\author{Jon P. Fulbright\altaffilmark{1} and Robert P. Kraft\altaffilmark{2}}

\affil{UCO/Lick Observatory, Dept of Astronomy, University of California,
    Santa Cruz, CA 95064}
\altaffiltext{1}{Email:  jfulb@ucolick.org}
\altaffiltext{2}{Email:  kraft@ucolick.org}

\begin{abstract}
\singlespace

      Recent LTE analyses (\cite{i98}, \cite{b99}) of
the OH bands in the optical ultraviolet spectra of nearby metal-poor subdwarfs
indicate that oxygen abundances are generally higher than those previously
determined from the analysis of the [O I] doublet in the spectra of 
low-metallicity giants. On average, the difference increases with decreasing
metallicity and reaches $\Delta$[O/Fe] $\sim$ +0.6 dex as [Fe/H] 
approaches -3.0. 
Employing high resolution (R = 50000), high S/N ($\sim$ 250) echelle spectra
of the two stars found by Israelian et al. (1998) to have the highest 
[O/Fe]-ratios,
viz, BD +23 3130 and BD +37 1458, we conducted abundance analyses based
on about 60 Fe I and 7-9 Fe II lines. We determined from Kurucz LTE
models the appropriate values of $T_{eff}$, log g, [Fe/H], $v_t$, as well as
abundances of Na, Ni, and the traditional alpha-elements Mg, Si, Ca and Ti,
independent of the calibration of color vs $T_{eff}$ scales. We determined 
oxygen
abundances from spectral synthesis of the stronger line ($\lambda$ 6300 \AA)
of the [O I] doublet. The ionization equilibrium of Fe indicates that
these two stars are subgiants, rather than subdwarfs, and our derived
values of $T_{eff}$ are 150 to 300K lower than those assumed by the previous 
investigators, although the resulting [Fe/H]-values differ only slightly. The 
syntheses of the [O I] line lead to smaller values of [O/Fe], consistent with 
those found earlier among halo field and globular cluster giants. We obtain
[O/Fe] = +0.35 $\pm$ 0.2 for BD +23 3130 and +0.50 $\pm$ 0.2 for BD +37 1458.
In the former, the [O I] line is very weak ($\sim$ 1 m\AA), so that the
quoted [O/Fe] value may in reality be an upper limit.
Therefore in these two stars a discrepancy exists between the [O/Fe]-
ratios derived from [O I] and the OH feature, and the origin of this
difference remains unclear. Until the matter is clarified, we suggest it is
premature to conclude that the ab initio oxygen abundances of old,
metal-poor stars need to be revised drastically upward.

\end{abstract}


\section{Introduction}

\singlespace
      After hydrogen and helium, oxygen is the most abundant element in the
Universe. It is well-known that knowledge of [O/Fe]-ratios in old, metal-poor 
stars is required if one is to test theories of Galactic chemical evolution. 
In addition, the oxygen abundance also affects both the energy generation and 
opacity of stars near the main sequence turnoff region of the color-magnitude
diagram, and thus affects the determination of ages of globular clusters and
the oldest stars (\cite{rc85}, \cite{v85}). The sensitivity of age
to [O/Fe] is not negligible: according to \cite{v92}, the gradient is 
$\Delta T/\Delta$ [O/Fe] = -4 Gyr/ 1 dex. Until recently, most determinations 
of 
[O/Fe] among old, metal-poor stars, have been confined to the analysis of the
forbidden [O I] doublet at $\lambda\lambda$ 6300, 6364 \AA. These transitions 
arise from the ground state and appear in absorption in the sun and also in 
metal-poor giants. All such studies agree (cf. reviews by \cite{sun93},
\cite{k94}, \cite{b94}, \cite{m97}) that [O/Fe] increases from 
$\sim0.0$ at solar metallicity to a value near +0.4 at [Fe/H] $\sim$ -1.0, 
after which [O/Fe] levels off near this last-named value and is essentially 
constant (with some scatter) to metal-poor values as low as [Fe/H] $\sim$ -3. 
The behavior of oxygen therefore mimics that of the other so-called alpha 
elements, e.g., Mg, Si and Ca. The only exception seems to be the behavior 
of [O/Fe] among globular
cluster giants where, presumably as a result of deep mixing of the stellar
envelope through the CNO hydrogen-burning shell, oxygen is significantly
depleted, sometimes well below the value [O/Fe] $\sim$ 0.0, and transmuted to
nitrogen (e.g., \cite{k97}).

      Apparently oxygen depletion among field halo giants, similar to that
experienced by globular cluster giants, does not take place
(\cite{h98}). However, since main sequence stars do not mix their
surface layers into regions where oxygen depletion could occur, one naturally
assumes that oxygen abundances determined from subdwarfs would be more secure
than those determined from their giant star analogues. Unfortunately, it is 
impractical to use the [O I] doublet to determine [O/Fe] among metal-poor
subdwarfs. Owing to that fact that these stars are generally hotter, 
have higher atmospheric density and have  
much decreased ratio of line to continuous opacity in comparison with giants, 
the $\lambda\lambda$ 6300, 6364 \AA$\;$ lines virtually disappear from their 
spectra. The solar EW of $\lambda$ 6300 \AA$\;$, the stronger of the two 
forbidden lines, is 6 m\AA.
In a giant of metallicity [Fe/H] = -2.2, for example, the 
EW(6300) $\sim$ 30 m\AA$\;$
when [O/Fe] = +0.4 (e.g., \cite{s97}). We calculated model synthetic
spectra of the $\lambda$ 6300 \AA$\;$ region in a subdwarf having solar values of 
$T_{eff}$
and log g and also having [Fe/H] = -2.2 and found that EW(6300) = 1 m\AA, even
for an [O/Fe]-ratio as large as +1.0. High resolution, high S/N analysis of the
two well-known subdwarfs HD 74000 and HD 25329 (\cite{bs94}) 
confirms this surmise. Clearly the discrimination of [O/Fe]-values among very
metal-poor subdwarfs in the [O/Fe]-range 0.0 to +1.0 is beyond present 
observational capabilities, if one employs only the forbidden [O I] doublet.

     The IR permitted O I triplet lines centered near $\lambda$ 7774 \AA$\;$ 
provide
an alternative means of estimating [O/Fe], their EWs being in the 10-20 m\AA$\;$
range even for very metal-poor subdwarfs having $T_{eff}$ and log g similar
to that of the sun (\cite{ar89}). Analysis of these lines has in
general led to values of [O/Fe] several tenths dex higher than those derived
from the [O I] lines in metal-poor giants (\cite{ar89}, \cite{t92},
\cite{c97}). The discrepancy increases with decreasing  
metallicity and amounts to more than 0.5 dex as [Fe/H] approaches -3.
Since the triplet lines arise from a level more than 9 eV above the ground
state of the oxygen atom, \cite{k93} suggested that an increase in the adopted
$T_{eff}$ scale (Carney 1983) of subdwarfs by 150-200K would resolve the 
problem.  This solution was criticized by \cite{bc96}, who, in analyzing the 
infrared OH and CO bands of the well-known, intermediately metal-poor 
([Fe/H] = -1.22) subdwarf HD 103095, determined the low oxygen abundance 
of [O/Fe] = +0.29, consistent with the originally adopted \cite{c83} 
$T_{eff}$ scale.
Other approaches to the problem involving radiative transfer (\cite{kn95})
and chromospheric heating (\cite{t95}) effects have also been 
invoked, but a generally agreed upon solution has not be forthcoming.

      Analysis of the low-excitation bands of OH, which lie in the (optical)
UV just longward of the atmospheric cutoff, provides a third means of
obtaining [O/Fe]-ratios (\cite{bhc84}), but the procedure is
rendered difficult because of flux limitations and severe line blending even
in metal-poor stars. \cite{bsr91} and \cite{n94} derived
[O/Fe]-values for a few metal-poor stars, and concluded that these were
in agreement with values derived from analysis of the [O I] lines. Recently,
however, an analysis of 23 stars, mostly low-metallicity subdwarfs, based on
the UV OH features, has led to the conclusion (\cite{i98}, hereafter
I98) that oxygen is strongly overabundant, the [O/Fe]-ratio increasing from
$\sim$+0.6 to $\sim$+1.0 as [Fe/H] decreased from -1.5 to -3.0. These results 
are in close agreement with the results obtained earlier from the IR permitted
oxygen triplet. A similar conclusion has been reached from a study of 24
(mostly) unevolved low-metallicity stars (\cite{b99}, hereafter B99),
again based on the UV OH bands. If these higher [O/Fe]-values are indeed
correct, three conclusions emerge, if at the same time, the smaller [O/Fe]-
values derived for giants from analysis of the [O I] lines are also correct.
First, all metal-poor giants must indeed have undergone deep mixing which
has permitted the ashes of O to N processing to reach the surface. This would,
of necessity, be true for halo giants, even those not members of globular 
clusters. Moreover, even those globular cluster giants having [O/Fe] near
+0.4 must themselves also have undergone deep mixing. Second, oxygen does not 
share the same degree of modest ``overabundance" as do other alpha elements such
as Mg, Si and Ca, all of which are predicted to arise from the same 
nucleosynthetic processes in Type II supernovae (e.g. \cite{ar95}, Fig 1). 
Finally, the ages of globular clusters are presumably 1 to 2 Gyr
younger than one would have inferred from the [O I]-based oxygen abundances
(\cite{v85}, \cite{ch98}).

     It is also possible, of course, that the disagreement between the
oxygen abundances arises not because of some real physical effect, but rather
from some inadequacy of the analysis or modelling procedure. We have already
noted that, among low-metallicity subdwarfs, the [O I] lines become too
weak to provide useful discriminatory input when [O/Fe] $<$ +1.0. However,
two stars in the list of I98, BD +23 3130 and BD +37 1458, are actually
subgiants: their spectra, as we shall show, contain measureable lines of
the [O I] doublet. These two stars are also the most overabundant in oxygen
of all stars in the I98 list, with [O/Fe] values, derived from the UV
OH bands, of +1.17 and +0.97, respectively. The star BD +37 1458 is also
to be found in the list by B99, for which they derive a similarly large
oxygen abundance of [O/Fe] = +0.83 (King 1993 $T_{eff}$ scale). 

     The remainder of this paper concerns the analysis of [O/Fe] for these two
subgiants, making use of observations of the $\lambda$ 6300 [O I] line derived 
from high resolution, high S/N spectra. We conduct a full LTE analysis of a 
large sample of Fe I and Fe II lines, from which we determine $T_{eff}$ 
and log g directly from Kurucz model atmospheres, without appealing to 
temperature scales derived from colors. This procedure leads to oxygen 
abundances that do not agree with those derived either from the IR triplet
permitted lines or the UV OH bands. Instead, we derive [O/Fe]-values that
are consistent with those derived from analysis of [O I]-doublet in
bright, low-metallicity field giants.

\section{Observations and Reductions}

     The two stars selected for observation, BD +23 3130 and BD +37 1458, were
taken from the lists of metal-poor stars analyzed by I98 and B99 because we
inferred that their surface gravities corresponded more closely to subgiants
than subdwarfs. As such, we suspected that they would have measureable [O I]
lines at $\lambda$ 6300 \AA. Both stars 
were observed on two different nights using the Hamilton Echelle Spectrograph
(\cite{v87}), operated at the coude focus of the Shane 3-m Telescope
of Lick Observatory.  The spectrograph was set at a (two pixel) spectral 
resolution of R = 50,000. The CCD detector was a thinned, cosmetically 
excellent, LL2-LA6-0 chip with dimensions of 2048 x 2048 pixels. 
Observational details are listed in Table 1.

     The frames were reduced using standard IRAF routines. First the overscan
pedestal was removed, followed by the bias. A normalized flat field was
created from several quartz exposures, and used to correct the pixel-to-
pixel variations. Occasional bad columns were corrected by interpolation.
After apertures were defined and traced, the frames were corrected for
scattered light and the spectra extracted. Finally, a wavelength solution,
created from Th/A lamp exposures, was attached to the extracted spectra.

     During both nights of observation, extremely high S/N spectra were
taken of hot, rapidly rotating stars. These were used to remove
the atmospheric $O_2$ lines present in the $\lambda$ 6300 \AA$\;$ spectral 
region.  Following the procedures outlined in detail elsewhere (e.g., 
\cite{matt96}), the hot star spectra were adjusted in strength so that 
the $O_2$ lines matched those of the program star; the adjusted hot star 
spectrum was then divided from the program star. Fortunately the [O I] line 
did not lie near an $O_2$ feature in either star, but 
the $\lambda\lambda$ 6297 \AA$\;$ and 6301 \AA$\;$ 
Fe I lines of BD +23 3130 were blended with atmospheric $O_2$ lines, as was the
$\lambda$ 6301 \AA$\;$ Fe I line of BD +37 1458. Thus great care was taken in
matching the $O_2$ features of the hot star with those of the program objects.
None of these Fe I lines were used to determine the atmospheric parameters
of either program star. The corrected spectrum of this region is, however,
shown in the syntheses discussed later (see Section 5), in which the
flat fielding was accomplished using the hot stars themselves, without
using the quartz exposures. Experience shows (e.g., \cite{s91})
that hot-star flatfields are superior to quartz flatfields as a basis
for testing synthetic spectra, especially of weak lines, possibly
because the hot star beam fills the collimator of the spectrograph in 
the same manner as does the beam of the program star. Finally, we estimate
that the resulting program star spectra have a S/N $\sim$ 250 in the vicinity
of the $\lambda$ 6300 feature.

\section{Atomic Data and Equivalent Width Measurements}     

     The atomic data for the Fe I lines used here were taken from the
work of the Oxford group (\cite{bl86}) and from \cite{o91}.
Only lines with gf-values certain to 10\% or better were adopted
from the latter source. In cases where a line was listed in both of the
sources, we found that the agreement was generally good, but we adopted
the Oxford values in the end. The atomic data for the Fe II lines come
from \cite{b91} and \cite{kk87} (see also the
discussion in \cite{s91}). The atomic data for the [O I] doublet
and the O I permitted triplet lines are well known (cf. \cite{l78}).
The gf-values for other elements (see Section 6) follow the recommendations
given in \cite{s94}, Shetrone (1996) and Kraft et al. (1997), to
which the reader if referred for details.

     The EWs of the lines were measured using basic IRAF routines. In
most cases, we fitted a Gaussian profile to each line, but in the few
instances in which a Gaussian fit proved to be poor, we measured the line
by direct integration under the continuum. The two methods gave consistent
results when tested on lines for which the Gaussian fit was satisfactory.
Lines that were badly blended or located on (occasional) bad columns of
the CCD chip were not measured.  In the spectrum of BD +37 1458, we 
measured 69 lines of Fe I and 9 lines of Fe II, ranging in EW from 4 to 100 
m\AA.
In the spectrum of the more metal-poor BD +23 3130, we were able to measure 56 
Fe I and 7 Fe II lines, over the same range of EW. The S/N proved high enough 
that measurements down to the 2-3 m\AA$\;$ level can be considered reliable. 
Table 2 lists the adopted atomic parameters and EWs of all the lines used in this study.

\section{Stellar Parameters and Fe Abundances}

     Analysis of the EWs was done iteratively using the LTE code MOOG
(\cite{s73}) together with \cite{k92} model atmospheres. The latter
were also used by both I98 and B99. Trial values for the stellar parameters
($T_{eff}$, log g and $v_t$) were those of I98. The method of
analysis closely followed that of the ``Lick/Texas" group, and the reader
is referred to the last two of a series of papers by that group (\cite{k97},
\cite{s97}) for a more lengthy discussion of the procedures.
\footnote{We adopt a solar Fe-abundance of log epsilon(Fe) = 7.52. Otherwise
all solar abundances adopted here are those of \cite{ag89}.}

      The ab initio stellar parameters were adjusted so that (1) the
stronger Fe I lines gave the same abundance on the average as the weaker
lines, (2) the Fe I lines gave an Fe abundance that was on the average
independent of excitation potential of the Fe I lines, and (3) the
average abundance of Fe based on Fe II lines was the same as that of the
Fe I lines (to within 0.02 dex). Because the ratios [Mg/Fe] and [Si/Fe] among 
metal poor stars are generally positive (and near +0.4 to +0.5), and the 
solar abundances
of Fe, Mg and Si are comparable, the input model should allow for the increased
electron density coming from the ionization of Mg and Si. In prior 
investigations of the Lick/Texas group, that increase was simulated in the 
input models by adopting a value of [Fe/H] higher than the final iterated
value.  Thus, for example in the present case, in which the iterated final 
value of [Fe/H] for BD +37 1458 proved to be -2.31, we adopted an input model
abundance of [Fe/H] = -2.0.  Since, as proved finally to be the case, 
[Fe/H] = -2.3 whereas [Mg/Fe] and [Si/Fe] $\sim$ +0.5, a compromise input 
metallicity of -2.0 should allow for an ``average" increase in the free 
electron supply from Mg, Si and Fe by a factor of about 2 (i.e. factors 
of 3, 3, and 1 for Mg, Si and Fe, respectively).  The multiple iteration of
the basic parameters following the prescriptions of (1), (2) and (3) above,
led to the final adopted values listed in Table 3.
Plots of Fe abundance vs Fe I excitation
potential (sensitive to the adopted $T_{eff}$) and Fe abundance vs Fe I line-
strength (sensitive to microturbulent velocity $v_t$) are shown in Figures
1a for BD +21 3130 and 1b for BD +37 1458 for the parameters adopted here.

As a check on the reliability of our procedure in which we simulate the effect 
on the models of an ``overabundance" of Mg and Si relative to Fe by 
substituting an input model with a ``compromise" increase in all three 
elements, we considered several LTE Kurucz models kindly supplied by 
Bruce Carney in which actual overabundances of Mg and Si by 0.5 dex relative 
to Fe were directly incorporated.  For example, we compared the effect of 
the ``simulated" ($T_{eff} = 5100$ K, 
log g = 2.9, [Fe/H] = -2.0, [Mg/Fe] = [Si/Fe] = 0.0) and ``real" 
($T_{eff} = 5100$ K, log g = 2.9, [Fe/H] = -2.31, [Mg/Fe] = [Si/Fe] = +0.5)
model on the final iterated Fe and O abundances of BD +37 1458, in which
the similated values led to [Fe/H] = -2.31.  We find that the changes
are extremely minor, amounting at most to 0.01 dex in [Fe/H] and 0.05 dex in
[O/Fe].  The reason these changes are minor is no doubt a result of the small
modifications induced in the free electron supply, since at each reference 
depth in the continuous opacity $\tau_{5000}$, the change in $T_{eff}$ and
$N_e$ amounts to no more than 20 K and 7 percent at the level of the continuum,
respectively.  
This in turn reflects the fact that for the overabundances of Fe, Mg and Si
dicussed above, 60 percent or more of the free electron supply is derived
from the ionization of H at every optical depth in the atmosphere, and thus
is virtually independent of the precise value of the input abundances of 
these elements.

     In Table 3, we list our final values of the stellar parameters along
with those adopted by I98 and B99. We obtain values of $T_{eff}$, derived
from our LTE analysis of roughly 60 Fe I lines, that are considerably lower
(by 150 to 350K) than those of the other investigations. For comparison
purposes, we plot Fe abundance vs Fe I excitation potential and vs Fe I line 
strength, using our measured Fe I EWs, for the values of $T_{eff}$ and 
$v_t$ adopted by
I98 and B99 (Figures 2a-d). For the B99 paper, the plot corresponds
to the adoption of the \cite{k93} $T_{eff}$ scale. Inspection of these Figures
reveals that adoption of the I98 and B99 values of $T_{eff}$ for these two 
stars leads to unacceptably large slopes and scatter in both the excitation
and turbulence plots. Compared with the other two investigations, our
values of $T_{eff}$ and log g are lower, but the differences in 
metallicity, i.e., [Fe/H], are fairly small ($\leq$ 0.2 dex). 

     As noted above, we obtain log g using the requirement that the Fe
abundance derived from the Fe II lines shall closely equal that derived from
the Fe I lines. In Table 3, we list the values of log $\epsilon$(Fe I) and
log $\epsilon$(Fe II) from our model parameters for the two stars, and it 
will be
seen that this condition is met to within 0.01 dex. We derive also (Table 3) the
mean abundances of Fe I and Fe II for the I98 and B99 model parameters, making
use of our EW measurements. These lead to discrepancies in the mean abundance of
Fe from Fe I and Fe II of 0.1 to 0.2 dex. We note in the next section that
the value of [O/Fe] is somewhat sensitive to the choice of log g. Thus, as
an experiment, we examined the Fe ionization equilibrium for an arbitrary
0.3 dex change in log g, having fixed our value of $T_{eff}$ in accordance 
with the
Fe I excitation plot exhibited in Figures 1a and 1b. The changes in log
$\epsilon$(Fe) from Fe I and Fe II lines are listed in Table 3, and lead to
discrepancies in Fe abundance somewhat more than 0.1 dex.

     Generally it has been found (cf. \cite{k97}, \cite{s97},
\cite{gw98}) that, among globular cluster giants, estimates
of surface gravity based on the ionization equilibrium of Fe I and Fe II
are in excellent agreement with estimates based on values of the effective
temperature, mass and luminosity expected for such stars. Thus, having settled
on values of $T_{eff}$ for BD +23 3130 and BD +37 1458, we can check whether
our corresponding log g's are consistent with expectation based on stellar
evolution, using the color-magnitude arrays of globular clusters. BD +37 1458,
for example, has a metallicity ([Fe/H] = -2.3) which is close to that of
M92 (\cite{zw84}, \cite{s91}, \cite{matt96}). Theoretical models
of a cluster with this metallicity require log g = 2.7 if $T_{eff}$ = 5100K 
and the
age is 14 Gyr (\cite{sc91}); the models of \cite{c82}
for M92 yield log g = 2.8 for this $T_{eff}$. Our value of log g = 2.9 
seems about 
right. If this star were actually located in M92 it would have $M_{Vo}$ = +1.9
for an assumed distance modulus of $(m-M)_o$ = 14.49 (\cite{d93}) and 
$(B-V)_o$ = +0.62 (\cite{sh88}). The observed $(B-V)$ = +0.60 
(\cite{clla}).
Interstellar reddening in this direction appears to be zero
or quite small even though b = 9.8 deg: for five stars selected from the
Bright Star Catalog (\cite{bsc}) within 5 deg of BD +37 1458 and having 
$b \leq 10$ deg, $<E(B-V)>$ = +0.02 $\pm$ 0.03 ($\sigma$). Thus the observed 
and ``predicted"
\footnote{From the same source, one finds $<E(B-V)>$ = +0.01 $\pm$ 0.13 for the
five stars nearest to BD +37 1458, and having D $>$ 100 pc.  Stars fainter than
those in the Bright Star Catalog, and close to BD +37 1458, are found in
the \cite{hip}.  There are 3 stars (a K0 giant, a K0 supergiant,
and a main-sequence A2 star) between $l = 171^o$ and $175^o$, having distances 
greater than 150 pc and $b < 8^o$.  For these $<E(B-V)> = 0.01 \pm 0.03$. 
Unfortunately the spectral types are those assigned by the Draper Catalog; MK
classifications are unknown.  In the direction of BD +37 1458, the dust 
reddening maps (\cite{sfd98}) lead to E(B-V) = 0.28.  Evidently some
dust clouds are located in a region beyond BD +37 1458.}
M92 colors are quite close. The distance to BD +37 1458 thus becomes 256 pc,
which is only slightly beyond the upper limit of the Hipparcos distance
plus its estimated error: 173 (+60, -32) pc. The I98 model is also consistent
with ``membership" in M92, the increase in $T_{eff}$ (from 5100K to 5260K) being
compensated by the corresponding increase in log g (from 2.8 to 3.0), and
the $M_{Vo}$ derived leads to a distance lying within the error bars of the
Hipparcos-based value. The values of $T_{eff}$ and log g assigned to this 
star by B99 are
still larger than the I98 values and again are consistent with the color-
magnitude array of M92. Identification of BD +37 1458 as a star consistent
with membership in M92 therefore does not help to discriminate which of the
models discussed here is correct. But all models do place the star above the
main sequence in the subgiant domain.

     There are no galactic globular clusters with metallicities as low as
[Fe/H] = -2.8, so a direct comparison of BD +23 3130 with globular cluster stars
cannot be made. There appear also to be no available theoretical isochrones
corresponding to clusters with metallicities below that of M92. Examination
of the Straniero and Chieffi isochrones with respect to [Fe/H] suggests that
subgiant and giant branches crowd more closely together as metallicity 
decreases; thus extrapolation toward a metallicity as low as [Fe/H] = -2.8 is a
dangerous procedure. Nevertheless, we estimate that, relative to BD +37 1458,
BD +23 3130 should be $\sim$0.05 mag bluer in (B-V) because of its reduced 
metallicity and $\sim$0.10 mag redder because of its reduced $T_{eff}$, and 
thus if
$(B-V)_o$ = 0.60 for the former star, $(B-V)_o$ = 0.65 for the latter. The
observed (B-V) = 0.64 (\cite{clla}). Once again we find $<E(B-V)>$ =
0.01 $\pm$ 0.07 from the five nearest stars in the Bright Star Catalog, so
interstellar reddening is quite small or vanishing. However, because
in color-magnitude diagrams, cluster giant branches are extremely steep,
we suggest that using estimates of color to derive values of $M_V$ is an unwise
procedure. The best we can say is that we expect BD +23 3130 to have a lower
log g than BD +37 1458, as we see from our spectroscopic determination.

     In Figures 3a and 3b, we synthesize three Fe I lines of similar 
intermediate excitation (2.4 to 2.6 eV) and differing strengths in the 
spectral 
region $\lambda$ 6134 \AA$\;$ to $\lambda$ 6140 \AA, using our parameters and 
those of I98 
and B99. The Hamilton spectra, binned by two pixels (= effective spectral 
resolution) are shown as dots. The parameters adopted by I98 and B99 
produce synthesized lines that are too weak compared with the observed spectra.
Obviously the discrepancies could be reduced simply by increasing the
Fe abundance, but such a change would introduce a serious discordance
between models and observations for lines of higher and lower excitation,
as seems clear from inspection of Figures 2a-d.

\section{Oxygen Abundances}

     Using the same techniques we employed to create Figures 3a and 3b, we
show in Figures 4a,b and 5a-d the spectral region surrounding the $\lambda$
6300 [O I] line in BD +23 3130 and BD +37 1458, respectively. Again our
observed Lick spectra (binned by 2) are shown as dots. In each figure, the Fe 
abundance was held constant while the O abundance was changed to create
plots (thin lines) for values of [O/Fe] = 0.0, +0.5, +1.0 and +1.5. The model
parameters and Fe abundances are those corresponding to each individual
investigation discussed in this paper.

     Turning first to the spectrum of BD +23 3130, we see that the present
synthesis gives a satisfactory representation of the Fe I lines at $\lambda
\lambda$ 6297.8 \AA, 6301.5 \AA$\;$ and 6302.5 \AA, and leads to an estimated 
oxygen abundance 
of [O/Fe] = +0.35 $\pm$ 0.2 (est. error, based on the S/N). 
However, the [O I] line is so weak (we estimate EW $\sim$ 1 m\AA) that the
quoted [O/Fe]-value may be an upper limit, in which case the true oxygen 
abundance is even smaller than [O/Fe] = +0.35.  The I98 model
produces Fe I lines that are much too weak, but corresponds to an estimated
oxygen abundance of [O/Fe] = +0.7 $\pm$ 0.2. Neither model comes close,
however, to the value [O/Fe] = +1.17 obtained by I98 from synthesis of the
OH features. Thus for this star, there is a serious discrepancy between the
oxygen abundances derived from [O I] and OH, even if we adopt the parameters
$T_{eff}$, log g, $v_t$ and [Fe/H] suggested by I98.

     The situation with respect to BD +37 1458 is less decisive. The synthesis
using the present parameters leads to a satisfactory representation of the
three Fe I lines listed in the preceding paragraph, and the corresponding 
oxygen abundance is [O/Fe] = +0.5 $\pm$ 0.2. The I98 model produces Fe I 
lines at $\lambda\lambda$ 6297.8 \AA$\;$ and 6301.5 \AA$\;$ that are much too weak, 
although
the modelling of the remaining Fe I line is satisfactory. The corresponding
oxygen abundance is [O/Fe] = +0.75 $\pm$ 0.2, a value reasonably close to
the value [O/Fe] = +0.97 obtained by I98 from the OH features. This star
was also considered by B99, who found [O/Fe] = 0.875 and 0.83, based on 
modelling the OH features, using the \cite{c83} and \cite{k93} $T_{eff}$
scales, respectively. These values of [O/Fe] are statistically indistiguish-
able from the value of [O/Fe] we derive from the [O I] line, using their
choices of the basic parameters. However, as we have seen, these models
give a poor representation of the Fe I lines (Figures 3a,3b), and
do not satisfy the requirements of the Fe I excitation and turbulence
plots (Figures 2a-d).

     We may also explore the changes in oxygen abundance derived from OH that
would be induced if I98 and B99 had adopted the values of $T_{eff}$ preferred
by our analysis.  According to B99, an increase in $T_{eff}$ by 100 K 
increases the derived O from OH by +0.16 dex.  But it also increases Fe by 
$\sim$ 0.07 dex.  So the net increase in [O/Fe] is about 0.09 dex.  
If I98 had adopted our $T_eff$ for BD +37 1458 their [O/Fe] would have gone
down by $\sim$ 0.15 dex.  In the case of BD +23 3130, the reduction would be
$\sim$ 0.25 dex.  

\section{Other Elements}

     In Table 4 and 5 we list, along with the determinations of [O/Fe] cited
in Section 5, abundances [el/Fe] for Na, the alpha elements Mg, Si, Ca
and Ti, and the Fe-peak element Ni for the two stars studied here. In both
Tables, the [el/Fe]-ratios are based on the EWs determined in this study.
The Tables differ in that the adopted Fe abundances of Table 4 are taken
as the Fe I abundances we derive using these EWs and the model parameters
($T_{eff}$, log g) used by I98 and B99, whereas the Fe abundances of Table 5
are those actually adopted by I98 and B99. We list also in these Tables
the mean [el/Fe]-ratios found by \cite{m95} for a large sample
of metal-poor stars having [Fe/H] $<$ -2.0.

     At any given [Fe/H], the scatter in [el/Fe]-ratios in the \cite{m95}
sample is typically $\sim$ 0.2 dex for stars more metal-rich than [Fe/H]
=-3.2. Thus we see from inspection of Tables 4 and 5 that almost all models
give [el/Fe]-ratios for BD +37 1458 and BD +23 3130 that are in reasonable
agreement with the mean [el/Fe]-ratios of the \cite{m95} sample. The
only significant departures seem to be those of the I98 models tabulated
in Table 5. For these the [el/Fe]-ratios appear to be rather large, and
the values for Na and Ni are particularly uncomfortable. However, this
departure is much reduced if we use the (mean) Fe-abundances we derive from
the present set of EWs and the I98 model parameters, rather than the 
Fe abundances actually adopted by I98.  Except for these two models of Table 5,
the [el/Fe] ratios for the ``$\alpha$ elements", Mg, Si and Ca are close
to their ``normal" expected values of $\sim$ +0.4 to +0.6 [cf., e.g., 
[$\alpha$/Fe]-ratios in M15 giants (Sneden et al. 1997).].

\section{Conclusions}

     In an analysis of the OH bands ($\lambda$ 3100 \AA) of 23 presumed 
subdwarfs, I98
found that the oxygen abundances were generally higher than the
traditionally accepted values that had been determined from the analysis
of the [O I] doublet in the spectra of low-metallicity giants. Similar
conclusions have also been reached by B99, again from analysis of the
OH bands in subdwarfs. The differences increase with decreasing metallicity,
and reach $\sim$+0.6 dex as [Fe/H] approaches -3.0. In particular, the two
highest [O/Fe]-ratios in the I98 sample were those found in BD +37 1458 and
BD + 23 3130, for which [O/Fe] = +0.97 and +1.17 for [Fe/H] = -2.40 and
-2.90, respectively. In their analysis of the former star, B99 found
a similar high value of [O/Fe] = 0.87 or 0.83, depending on the choice
of temperature scale. 

     Atmospheric modelling of very metal-poor ([Fe/H $<$ -2) subdwarfs shows 
that the [O I]-doublet, even for [O/Fe]-ratios as high as +1.0, practically
``disappears": the EW of the stronger member of the pair ($\lambda$ 6300A)
approaches 1.0 m\AA$\;$ or less. However, examination of the I98 sample 
indicates that
the two stars discussed above have surface gravities more like subgiants
than subdwarfs. These two stars thus presented the possibility that
the EWs of the $\lambda$ 6300 \AA$\;$ line of [O I] might be measureable in spectra
having sufficiently high S/N, and thus provide a check on the oxygen
abundances derived from the OH feature.

     Thus, on the basis of high resolution (R = 50000), high S/N ($\sim$250) 
echelle spectra of BD +23 3130 and BD +37 1458, we have determined abundances
[Fe/H], and [el/Fe]-ratios for Na, Ni, and the traditional alpha elements
Mg, Si, Ca and Ti. We conducted a full-scale abundance analysis based on the
EWs of about 60 Fe I and 7-9 Fe II lines, from which we determined from LTE
models the appropriate values of $T_{eff}$, log g, [Fe/H] and $v_t$, 
independent of
considerations based on the calibration of color scales. We determined 
oxygen abundances from synthetic spectra of the region encompassing the
[O I] line at $\lambda$ 6300 \AA. Our conclusions are these:

     (1) Based on the ionization equilibrium of Fe, we find that these two
stars are indeed subgiants, with surface gravities somewhat lower than those
adopted by I98 and B99.

     (2) Our values of $T_{eff}$ are also lower than those obtained by I98 and
B99, by amounts between $\sim$150 and 300K. Our [Fe/H] values are nevertheless
within $\sim$0.2 dex of those determined by I98 and B99. The metallicity of
BD +37 1458 is essentially the same as that of M92; all models are
in reasonable agreement with the assumption that this star is a surrogate
for an M92 star lying on the steep subgiant branch of that cluster. The
metallicity of +23 3130 is lower than that of any known galactic globular
cluster.

     (3) Except for two models listed in Table 5, all investigations
lead to plausible [el/Fe]-ratios for alpha elements, Na and Ni, when
compared with those found for metal-poor stars by \cite{m95}.

     (4) Our synthetic spectra of the region of the [O I] line at $\lambda$
6300 \AA$\;$ lead to the traditional oxygen abundances of [O/Fe] = +0.35 $\pm$ 
0.2 for BD +23 3130 and +0.50 $\pm$ 0.2 for BD +37 1458.  The [O/Fe]-value
quoted for BD +23 3130 may be an upper limit.

     (5) Use of our Fe I EWs leads to a fairly low value of [O/Fe] = +0.7 in
the case of BD +23 3130, even if we adopt the parameters of I98. For
the other star, an [O/Fe]-ratio closer to the value derived from the OH
feature is found if the parameters adopted by I98 and B99 are adopted.
These parameters, however, yield a poor representation of the Fe I
features in this star, as do the parameters adopted by I98 for BD +23 3130.

     (6) A discrepancy therefore remains among these two stars between
the [O/Fe]-ratio derived from [O I] and the OH feature. This is of 
some significance especially when one realizes that these two stars have 
the highest [O/Fe]-ratios among the stars considered by I98. The origin of
these differences remains unclear.

     (7) The procedures adopted here are identical with those previously
employed (e.g., \cite{s97}, \cite{k97}) in the analysis of the
spectra of globular cluster and halo field giants and subgiants. In these
studies, analysis of the Fe I and Fe II line spectra is a basic requirement
in the determination of atmospheric parameters. 

     (8) The results obtained here suggest that it is too early to conclude
that the oxygen abundances of old, metal-poor stars need to be revised
drastically upward. We suggest that input models for such stars need to
be examined in the light of a full LTE analysis of the Fe I and Fe II
spectrum, before values of $T_{eff}$, log g, $v_t$ and [Fe/H] are assigned.

\acknowledgments

     We are indebted to C. Sneden for use of his MOOG code, and have
benefitted from useful conversations with R. Peterson.  We are especially
grateful to Bruce Carney for supplying us with several $\alpha$-enhanced
Kurucz model atmospheres and for reading and commenting on a preliminary
version of this manuscript.  This research has
been supported by NSF Grant AST 96-18351, which we greatly appreciate.

\clearpage
\begin{deluxetable}{lcc}
\footnotesize
\tablenum{1}
\tablecaption{Log of Observations.}
\tablewidth{0pt}
\tablehead{
 & \colhead{BD +23 3130} & \colhead{BD +37 1458}}
\startdata
UT Date       & 8 Aug 1998 & 31 Oct 1998  \nl
Exposure      & 3600 s     & 3600 s       \nl
V             & 8.95       & 8.92         \nl
B-V           & 0.64       & 0.60         \nl
\enddata
\end{deluxetable}


\clearpage
\begin{deluxetable}{ r r r r r r}
\footnotesize
\tablenum{2}
\tablecaption{Line Measurements.}
\tablewidth{0pt}
\tablehead{
\colhead{Wavelength (\AA)} & \colhead{E.P. (eV)} & \colhead{log(gf)} & 
\colhead{EW (m\AA)} & \colhead{EW (m\AA)} & \colhead{Ref} \nl
 &  &  & \colhead{BD +37 1458} & \colhead{BD +23 3130} &}
\startdata
\multicolumn{1}{l}{Fe I}  & & & & & \nl
4531.15 & 1.49 & -2.155 & 58.2 & 44.5 & 1d\nl
4592.66 & 1.56 & -2.449 & 45.4 & 29.4 & 1d\nl
4602.01 & 1.61 & -3.154 & 12.1 &  7.6 & 1d\nl
4602.94 & 1.49 & -2.208 & 54.8 & 45.3 & 1d\nl
4643.46 & 3.64 & -1.147 &  7.5 &   -- & 2\nl
4647.43 & 2.94 & -1.351 & 23.6 & 13.4 & 2\nl
4733.60 & 1.49 & -2.987 & 24.3 & 14.0 & 1d\nl
4736.77 & 3.20 & -0.752 & 36.7 &   -- & 2\nl
4786.81 & 3.00 & -1.606 & 12.4 &  7.6 & 2\nl
4789.65 & 3.53 & -0.957 & 15.9 &   -- & 2\nl
4871.32 & 2.85 & -0.362 & 69.6 & 54.8 & 2\nl
4872.14 & 2.87 & -0.567 & 60.6 & 45.3 & 2\nl
4890.75 & 2.86 & -0.394 & 72.6 & 54.8 & 2\nl
4891.49 & 2.84 & -0.111 & 84.5 &   -- & 2\nl
4918.99 & 2.85 & -0.342 & 74.5 & 59.5 & 2\nl
4938.81 & 2.86 & -1.077 & 36.4 & 24.6 & 2\nl
4939.69 & 0.86 & -3.340 & 38.2 & 31.2 & 1b\nl
4994.13 & 0.91 & -3.080 & 48.8 & 38.5 & 1b\nl
5006.12 & 2.82 & -0.615 & 62.3 & 45.5 & 2\nl
5028.13 & 3.56 & -1.122 &  9.0 &   -- & 2\nl
5041.07 & 0.95 & -3.086 & 49.2 & 37.3 & 2\nl
5044.21 & 2.85 & -2.017 &  9.0 &   -- & 2\nl
5048.44 & 3.94 & -1.029 &  5.1 &   -- & 2\nl
5049.82 & 2.27 & -1.355 & 56.7 & 44.1 & 2\nl
5051.64 & 0.91 & -2.795 & 63.9 & 54.1 & 1b\nl
5068.77 & 2.93 & -1.041 &   -- & 22.2 & 2\nl
5079.23 & 2.20 & -2.067 & 30.6 & 20.1 & 1b\nl
5079.74 & 0.99 & -3.220 & 38.1 & 28.1 & 1b\nl
5083.34 & 0.96 & -2.958 & 52.6 & 42.4 & 1b\nl
5107.45 & 0.99 & -3.087 & 43.8 & 34.0 & 1b\nl
5107.65 & 1.56 & -2.418 & 44.6 & 33.8 & 1d\nl
5123.72 & 1.01 & -3.068 & 44.4 & 35.2 & 1b\nl
5127.36 & 0.91 & -3.307 & 37.3 & 30.6 & 1b\tablebreak
5151.92 & 1.01 & -3.322 & 32.2 & 22.1 & 1b\nl
5191.45 & 3.03 & -0.551 & 54.9 & 37.6 & 2\nl
5192.34 & 2.99 & -0.421 & 60.8 & 45.8 & 2\nl
5198.71 & 2.22 & -2.135 & 25.2 & 16.4 & 1c\nl
5216.28 & 1.61 & -2.150 & 56.9 & 43.8 & 1d\nl
5217.39 & 3.20 & -1.162 & 21.9 & 13.3 & 2\nl
5232.94 & 2.94 & -0.057 & 82.7 & 67.8 & 2\nl
5242.49 & 3.62 & -0.967 &   -- &  6.0 & 2\nl
5247.05 & 0.09 & -4.946 & 11.2 &   -- & 1a\nl
5288.53 & 3.68 & -1.508 &  3.9 &   -- & 2\nl
5307.37 & 1.61 & -2.987 & 18.6 & 11.4 & 1d\nl
5367.47 & 4.42 &  0.443 &   -- & 16.7 & 2\nl
5383.37 & 4.31 &  0.645 &   -- & 28.1 & 2\nl
5397.13 & 0.91 & -1.993 & 96.8 &   -- & 1b\nl
5405.78 & 0.99 & -1.844 & 98.3 &   -- & 1b\nl
5434.53 & 1.01 & -2.122 & 88.9 &   -- & 1b\nl
5501.46 & 0.95 & -3.046 & 51.2 & 40.1 & 2\nl
5506.78 & 0.99 & -2.797 & 60.9 & 51.0 & 1b\nl
5701.55 & 2.56 & -2.216 & 11.1 &  6.1 & 1f\nl
6065.49 & 2.61 & -1.530 & 35.5 & 22.9 & 1f\nl
6136.62 & 2.45 & -1.400 & 51.2 & 34.2 & 1e\nl
6137.70 & 2.59 & -1.403 & 45.2 & 30.0 & 1f\nl
6173.34 & 2.22 & -2.880 &  8.0 &   -- & 1e\nl
6219.29 & 2.20 & -2.433 & 16.5 &  9.9 & 1c\nl
6230.73 & 2.56 & -1.281 & 54.5 & 37.4 & 1f\nl
6246.32 & 3.59 & -0.877 & 19.8 &  9.9 & 2\nl
6252.56 & 2.40 & -1.687 & 38.9 & 28.7 & 1c\nl
6265.14 & 2.18 & -2.550 & 15.9 &  9.9 & 1e\nl
6297.80 & 2.22 & -2.740 & 11.0 &   -- & 1c\tablenotemark{a}\nl
6322.69 & 2.59 & -2.426 &  7.9 &   -- & 1f\nl
6358.69 & 0.86 & -4.468 &  7.6 &   -- & 1b\nl
6411.65 & 3.64 & -0.717 &   -- & 13.1 & 2\nl
6421.36 & 2.28 & -2.027 & 33.6 & 18.6 & 1e\nl
6430.85 & 2.18 & -2.006 & 39.4 & 25.2 & 1c\nl
6494.99 & 2.40 & -1.273 & 61.2 & 48.9 & 1e\tablebreak
6593.88 & 2.43 & -2.422 & 11.9 &  6.1 & 1c\nl
6677.99 & 2.68 & -1.418 & 42.3 & 27.5 & 2\nl
6945.21 & 2.42 & -2.482 &   -- &  6.7 & 1c\nl
6750.15 & 2.42 & -2.621 &  8.0 &   -- & 1e\nl
6945.21 & 2.42 & -2.482 & 11.4 &   -- & 1c\nl
6978.86 & 2.48 & -2.500 & 10.9 &   -- & 1c\nl
7511.02 & 4.16 &  0.099 & 32.9 & 19.7 & 2\nl
\multicolumn{1}{l}{Fe II}  & & & & & \nl
4555.88 & 2.83 & -2.290 & 33.2 & 28.7 & 3\nl
4576.34 & 2.84 & -2.822 & 13.4 & 10.6 & 4\nl
4582.83 & 2.84 & -3.094 & 10.9 &  6.2 & 4\nl
4620.52 & 2.83 & -3.079 &   -- &  5.7 & 4\nl
4923.92 & 2.89 & -1.240 & 77.0 &   -- & 3\nl
5197.58 & 3.23 & -2.233 & 25.6 & 19.6 & 4\nl
5234.63 & 3.22 & -2.151 & 28.5 & 23.0 & 4\nl
5316.62 & 3.15 & -1.850 & 44.9 & 35.7 & 3\nl
6247.56 & 3.89 & -2.329 &  6.2 &   -- & 4\nl
6456.39 & 3.90 & -2.075 & 10.1 &   -- & 4\nl
\multicolumn{1}{l}{Na I}  & & & & & \nl
5688.21 & 2.10 & -0.370 &   8.7 &    -- & 5\nl
5889.96 & 0.00 &  0.110 & 165.1 & 128.7 & 6\nl
5895.94 & 0.00 & -0.190 & 152.0 & 108.8 & 6\nl
\multicolumn{1}{l}{Mg I}  & & & & & \nl
4730.03 & 4.33 & -2.310 &   3.6 &    -- & 7\nl
5172.70 & 2.71 & -0.320 & 292.8 & 188.7 & 8\tablenotemark{a}\nl
5183.42 & 2.72 & -0.080 & 353.2 & 203.7 & 8\tablenotemark{a}\nl
5528.42 & 4.35 & -0.360 &  85.5 &  56.8 & 7\nl
5711.09 & 4.33 & -1.730 &  18.3 &   8.3 & 7\tablebreak
\multicolumn{1}{l}{Si I}  & & & & & \nl
5701.12 & 4.93 & -2.050 &  2.9 &  -- & 9\nl
5708.41 & 4.93 & -1.470 &   -- & 3.4 & 9\nl
5948.55 & 5.08 & -1.230 &  9.2 & 6.5 & 9\nl
7415.96 & 5.61 & -0.710 & 11.5 & 5.3 & 9\nl
7423.52 & 5.61 & -0.780 & 13.4 & 7.7 & 9\nl
\multicolumn{1}{l}{Ca I}  & & & & & \nl
4578.56 & 2.52 & -0.700 & 13.5 &  6.6 & 10\nl
5261.71 & 2.52 & -0.580 & 18.8 &  9.9 & 10\nl
5349.47 & 2.71 & -0.310 & 20.2 & 11.2 & 10\nl
5581.97 & 2.52 & -0.560 & 25.0 &  9.5 & 10\nl
5588.76 & 2.52 &  0.360 & 62.8 & 37.4 & 10\nl
5590.12 & 2.52 & -0.570 & 18.7 &  9.7 & 10\nl
5857.46 & 2.93 &  0.240 & 35.3 & 21.8 & 10\nl
6122.23 & 1.89 & -0.320 & 72.0 & 46.8 & 11\nl
6161.30 & 2.52 & -1.270 &  4.2 &   -- & 10\nl
6162.18 & 1.90 & -0.090 & 87.5 & 61.0 & 11\nl
6163.75 & 2.52 & -1.290 &  7.5 &   -- & 10\nl
6166.44 & 2.52 & -1.200 & 10.6 &   -- & 10\nl
6169.04 & 2.52 & -0.800 & 15.6 &  8.9 & 10\nl
6169.56 & 2.52 & -0.480 & 21.5 & 12.3 & 10\nl
6439.08 & 2.52 &  0.390 & 66.7 & 42.8 & 10\nl
6455.60 & 2.52 & -1.290 &  6.1 &   -- & 10\nl
6499.65 & 2.52 & -0.820 & 15.0 &   -- & 10\nl
\multicolumn{1}{l}{Ti I}  & & & & & \nl
4555.49 & 0.85 & -0.490 &  8.0 &  5.5 & 12\tablenotemark{a}\nl
4617.28 & 1.75 &  0.390 &  7.7 &  5.6 & 12\nl
4623.10 & 1.74 &  0.110 &  4.7 &   -- & 12\nl
4681.92 & 0.05 & -1.070 & 18.7 & 10.5 & 12\nl
4840.88 & 0.90 & -0.510 & 10.2 &  5.6 & 12\nl
4885.09 & 1.89 &  0.360 &  5.0 &   -- & 12\nl
4991.07 & 0.84 &  0.380 & 43.6 & 32.5 & 12\nl
4999.51 & 0.83 &  0.250 & 38.2 & 24.8 & 12\tablebreak
5020.03 & 0.84 & -0.410 & 13.7 &  8.5 & 12\nl
5024.85 & 0.82 & -0.600 &  9.5 &  5.6 & 12\nl
5064.66 & 0.05 & -0.990 & 22.8 & 15.5 & 12\nl
5173.75 & 0.00 & -1.120 & 19.0 & 11.4 & 12\tablenotemark{a}\nl
5210.39 & 0.05 & -0.880 & 27.4 & 17.5 & 12\nl
6258.71 & 1.46 & -0.270 &  5.0 &   -- & 12\nl
6261.11 & 1.43 & -0.480 &  4.0 &   -- & 12\nl
\multicolumn{1}{l}{Ti II}  & & & & & \nl
4708.67 & 1.24 & -2.210 & 13.5 &  9.5 & 12\nl
5154.08 & 1.57 & -1.920 & 22.1 & 16.5 & 12\nl
5185.91 & 1.89 & -1.350 & 21.2 & 18.6 & 12\nl
5226.54 & 1.57 & -1.300 & 47.0 & 33.3 & 12\nl
5336.79 & 1.58 & -1.700 & 26.9 & 19.6 & 12\nl
\multicolumn{1}{l}{Ni I}  & & & & & \nl
4756.52 & 3.48 & -0.340 &   -- &  5.3 & 13\nl
4829.03 & 3.54 & -0.330 &  7.1 &   -- & 13\nl
4831.18 & 3.61 & -0.420 &  6.2 &   -- & 13\nl
4904.42 & 3.54 & -0.170 &   -- &  5.3 & 13\nl
5017.58 & 3.54 & -0.080 & 10.2 &  6.7 & 13\nl
5081.12 & 3.85 &  0.300 & 14.6 &  7.5 & 13\nl
5476.92 & 1.83 & -0.890 & 60.3 & 46.9 & 13\nl
5754.67 & 1.93 & -2.330 &  6.4 &   -- & 13\nl
6643.65 & 1.68 & -2.300 & 10.6 &  6.7 & 13\nl
6767.78 & 1.83 & -2.170 & 12.1 &  6.8 & 13\nl
\tablenotetext{a}{Line not used in abundance analysis.}
\tablerefs{
 (1) ``Oxford" group:  (a) Blackwell et al. 1979a, (b) Blackwell et al. 1979b,
     (c) Blackwell et al. 1982a, (d) Blackwell et al. 1980, 
     (e) Blackwell et al. 1982b, (f) Blackwell et al. 1982c;
 (2) O'Brian et al. 1991;
 (3) Kroll \& Kock 1987;
 (4) Biemont et al. 1991;
 (5) Lambert \& Warner 1968;
 (6) Weise et al. 1969;
 (7) Fuhrmann et al. 1995;
 (8) Thevenin 1989;
 (9) Garz 1973;
(10) Smith \& Raggett 1981;
(11) Weise \& Martin 1980;
(12) Martin et al. 1988;
(13) Fuhr et al. 1988. }
\enddata
\end{deluxetable}

\clearpage
\begin{deluxetable}{lcccccccc}
\footnotesize
\tablenum{3}
\tablecaption{Atmospheric Parameters.}
\tablewidth{0pt}
\tablehead{
 & \colhead{$T_{eff}$} & \colhead{$\log g$} & \colhead{[Fe/H]} & \colhead{$v_t$}
 & \colhead{$\log \epsilon$(Fe I)} & \colhead{$\sigma$(Fe I)} &  
 \colhead{$\log \epsilon$(Fe II)} & \colhead{$\sigma$(Fe II)}}

\startdata
BD +23 3130 \nl
This Paper             &4850 K&2.00& -2.84 & 1.35 & 4.69 & 0.04 & 4.68 & 0.08\nl
This Paper, log g down &4850 K&1.70& -2.84 & 1.35 & 4.70 & 0.04 & 4.58 & 0.08\nl
This Paper, log g up   &4850 K&2.30& -2.84 & 1.35 & 4.67 & 0.04 & 4.78 & 0.08\nl
I98                    &5130 K&2.50& -2.90 & 1.00 & 5.02 & 0.09 & 4.89 & 0.08\nl
\hline
BD +37 1458 \nl
\nl
This Paper             &5100 K&2.90& -2.31 & 1.40 & 5.21 & 0.04 & 5.21 & 0.08\nl
This Paper, log g down &5100 K&2.60& -2.31 & 1.40 & 5.22 & 0.04 & 5.11 & 0.07\nl
This Paper, log g up   &5100 K&3.20& -2.31 & 1.40 & 5.20 & 0.04 & 5.32 & 0.08\nl
I98                    &5260 K&3.00& -2.40 & 1.00 & 5.47 & 0.10 & 5.30 & 0.08\nl
B99, K93 scale         &5554 K&3.62& -2.06 & 1.50 & 5.64 & 0.09 & 5.48 & 0.07\nl
B99, C83 scale         &5408 K&3.41& -2.14 & 1.50 & 5.50 & 0.07 & 5.40 & 0.08\nl

\enddata
\end{deluxetable}

\clearpage
\begin{deluxetable}{l|cc|cccc|c}
\footnotesize
\tablenum{4}
\tablecaption{Derived Abundances.}
\tablewidth{0pt}
\tablehead{
 \colhead{} & \multicolumn{2}{|c|}{BD +23 3130} & 
  \multicolumn{4}{c|}{BD +37 1458} & \colhead{McWilliam} \nl
 \colhead{} & \multicolumn{1}{|c}{This Paper} & 
  \multicolumn{1}{c|}{$\;\;$I98$\;\;$} & \multicolumn{1}{c}{This Paper} & 
  \multicolumn{1}{c}{$\;\;$I98$\;\;$} & \multicolumn{1}{c}{B99-K93} & 
  \multicolumn{1}{c|}{B99-C83} & \colhead{et al. 1995} }
\startdata
log $\epsilon$(Fe I) &  4.69 &  5.02 &  5.21 &  5.47 &  5.64 &  5.50 & \\
$[Fe/H]_{Fe I}$  & -2.83$\;$ & -2.50$\;$ & -2.31$\;$ & -2.05$\;$ & -1.88$\;$ & -2.02$\;$ &\\
$\sigma$         &  0.04 &  0.09 &  0.04 &  0.10 &  0.09 &  0.07 & \\
\hline
log $\epsilon$(Fe II)&  4.68 &  4.89 &  5.21 &  5.30 &  5.48 &  5.40 & \\
$[Fe/H]_{Fe II}$ & -2.84$\;$ & -2.63$\;$ & -2.31$\;$ & -2.22$\;$ & -2.04$\;$ & -2.12$\;$ & \\
$\sigma$         &  0.08 &  0.08 &  0.08 &  0.08 &  0.07 &  0.08 & \\
\hline
log $\epsilon$([O I])&  6.45 &  6.73 &  7.12 &  7.28 &  7.67 &  7.49 & \\
$[O/Fe]$         &  0.35 &  0.30 &  0.50 &  0.40 &  0.62 &  0.58 & \\
$\sigma$         &  0.20 &  0.20 &  0.20 &  0.20 &  0.20 &  0.20 & \\
\hline
log $\epsilon$(Na I) &  3.54 &  3.86 &  4.07 &  4.27 &  4.32 &  4.24 & \\
$[Na/Fe]$        &  0.05 &  0.04 &  0.06 &  0.00 & -0.12$\;$ & -0.06$\;$ &$\sim 0.1$ \\
$\sigma$         &  0.06 &  0.09 &  0.10 &  0.16 &  0.13 &  0.11 & \\
\hline
log $\epsilon$(Mg I) &  5.34 &  5.54 &  5.83 &  5.99 &  6.04 &  5.98 & \\
$[Mg/Fe]$        &  0.60 &  0.47 &  0.57 &  0.47 &  0.35 &  0.43 & 0.42 \\
$\sigma$         &  0.06 &  0.18 &  0.13 &  0.19 &  0.16 &  0.15 & \\
\hline
log $\epsilon$(Si I) &  5.43 &  5.55 &  5.87 &  5.94 &  6.05 &  6.00 & \\
$[Si/Fe]$        &  0.72 &  0.51 &  0.64 &  0.45 &  0.39 &  0.48 & 0.50 \\
$\sigma$         &  0.17 &  0.19 &  0.12 &  0.14 &  0.13 &  0.12 & \\
\hline
log $\epsilon$(Ca I) &  3.98 &  4.17 &  4.52 &  4.67 &  4.77 &  4.70 & \\
$[Ca/Fe]$        &  0.52 &  0.38 &  0.54 &  0.43 &  0.36 &  0.43 & 0.43\\
$\sigma$         &  0.09 &  0.12 &  0.12 &  0.16 &  0.14 &  0.13 & \\
\hline
log $\epsilon$(Ti) &  2.34 &  2.65 &  2.87 &  3.05 &  3.32 &  3.18 & \\
$[Ti/Fe]_{Ti}$     &  0.19 &  0.17 &  0.20 &  0.12 &  0.22 &  0.22 & 0.32\\
$\sigma$           &  0.10 &  0.12 &  0.09 &  0.13 &  0.12 &  0.11 & \\
\hline
log $\epsilon$(Ni I) &  3.46 &  3.72 &  3.96 &  4.13 &  4.34 &  4.22 & \\
$[Ni/Fe]$        &  0.05 & -0.02$\;$ &  0.03 & -0.06$\;$ & -0.02$\;$ &  0.00 & $\sim$0.0\\
$\sigma$         &  0.06 &  0.11 &  0.08 &  0.15 &  0.15 &  0.12 & \\

\enddata
\end{deluxetable}

\clearpage
\begin{deluxetable}{l|cc|cccc|c}
\footnotesize
\tablenum{5}
\tablecaption{Derived Abundances.}
\tablewidth{0pt}
\tablehead{
 \colhead{} & \multicolumn{2}{|c|}{BD +23 3130} & 
  \multicolumn{4}{c|}{BD +37 1458} & \colhead{McWilliam} \nl
 \colhead{} & \multicolumn{1}{|c}{This Paper} & 
  \multicolumn{1}{c|}{$\;\;$I98$\;\;$} & \multicolumn{1}{c}{This Paper} & 
  \multicolumn{1}{c}{$\;\;$I98$\;\;$} & \multicolumn{1}{c}{B99-K93} & 
  \multicolumn{1}{c|}{B99-C83} & \colhead{et al. 1995} }
\startdata
log $\epsilon$(Fe I) &  4.69 &  4.62 &  5.21 &  5.12 &  5.46 &  5.38 &\\
$[Fe/H]_{Fe I}$  & -2.83$\;$ & -2.90$\;$ & -2.31$\;$ & -2.40$\;$ & -2.06$\;$ & -2.14$\;$ &\\
\hline
log $\epsilon$(Fe II)&  4.68 &  4.62 &  5.21 &  5.12 &  5.46 &  5.38 &\\
$[Fe/H]_{Fe II}$ & -2.84$\;$ & -2.90$\;$ & -2.31$\;$ & -2.40$\;$ & -2.06$\;$ & -2.14$\;$ &\\
\hline
log $\epsilon$([O I])&  6.45 &  6.73 &  7.12 &  7.28 &  7.67 &  7.49 &\\
$[O/Fe]$         &  0.35 &  0.70 &  0.50 &  0.75 &  0.80 &  0.70 &\\
$\sigma$         &  0.20 &  0.20 &  0.20 &  0.20 &  0.20 &  0.20 &\\
\hline
log $\epsilon$(Na I) &  3.54 &  3.86 &  4.07 &  4.27 &  4.32 &  4.24 &\\
$[Na/Fe]$        &  0.05 &  0.44 &  0.06 &  0.35 &  0.06 &  0.06 & $\sim$0.1\\
$\sigma$         &  0.05 &  0.02 &  0.09 &  0.12 &  0.09 &  0.09 &\\
\hline
log $\epsilon$(Mg I) &  5.34 &  5.54 &  5.83 &  5.99 &  5.98 &  6.04 &\\
$[Mg/Fe]$        &  0.60 &  0.87 &  0.57 &  0.82 &  0.53 &  0.55 & 0.42\\
$\sigma$         &  0.05 &  0.16 &  0.12 &  0.16 &  0.13 &  0.13 &\\
\hline
log $\epsilon$(Si I) &  5.43 &  5.55 &  5.87 &  5.94 &  6.05 &  6.00 &\\
$[Si/Fe]$        &  0.72 &  0.91 &  0.64 &  0.80 &  0.57 &  0.60 & 0.50\\
$\sigma$         &  0.17 &  0.17 &  0.11 &  0.10 &  0.10 &  0.10 &\\
\hline
log $\epsilon$(Ca I) &  3.98 &  4.17 &  4.52 &  4.67 &  4.77 &  4.70 &\\
$[Ca/Fe]$        &  0.52 &  0.78 &  0.54 &  0.78 &  0.54 &  0.55 & 0.43\\
$\sigma$         &  0.08 &  0.08 &  0.11 &  0.13 &  0.11 &  0.11 &\\
\hline
log $\epsilon$(Ti) &  2.34 &  2.65 &  2.87 &  3.05 &  3.32 &  3.18 &\\
$[Ti/Fe]_{Ti}$     &  0.19 &  0.57 &  0.20 &  0.47 &  0.40 &  0.34 & 0.32\\
$\sigma$           &  0.09 &  0.07 &  0.08 &  0.09 &  0.09 &  0.08 &\\
\hline
log $\epsilon$(Ni I) &  3.46 &  3.72 &  3.96 &  4.13 &  4.34 &  4.22 &\\
$[Ni/Fe]$        &  0.05 &  0.38 &  0.03 &  0.29 &  0.16 &  0.12 & $\sim$0.0\\
$\sigma$         &  0.05 &  0.07 &  0.07 &  0.11 &  0.12 &  0.10 &\\

\enddata
\end{deluxetable}

\clearpage

\noindent {\bf Figure 1a}. Plots of abundance vs. excitation potential and line 
 strength for the individual Fe I lines of BD +23 3130 using the stellar 
 parameters adopted by this paper. 
 
\noindent {\bf Figure 1b}.  Same as Figure 1a, expect for BD +37 1458.  

\noindent {\bf Figure 2a}.  Same as Figure 1a, except using the parameters of 
 I98.
 
\noindent {\bf Figure 2b}.  Same as Figure 1b, except using the parameters of 
 I98.
 
\noindent {\bf Figure 2c}.  Same as Figure 1b, except using the parameters of  
 B99 (King 1993 scale).

\noindent {\bf Figure 2d}.  Same as Figure 1b, except using the parameters of
 B99 (Carney 1983 scale). 

\noindent {\bf Figure 3a}.  Observed and synthetic spectra of Fe I lines in the 
 $\lambda$ 6137 \AA$\;$ region of BD +23 3130.  The dots are the observed data 
 points (binned 2X), and the lines are the synthetic spectra 
 using the stellar parameters adopted by this paper (solid) 
 and by I98 (dotted). 

\noindent {\bf Figure 3b}.  Same as Figure 3a, except for BD +37 1458.  The lines are 
 the synthetic spectra for the stellar parameters adopted by this paper
 (solid), I98 (dotted), B99 using the King 1993 scale (short dash),
 and B99 using the Carney 1983 scale (long dash). 

\noindent {\bf Figure 4a}.  Observed and synthetic spectra for the $\lambda$ 
6300 \AA$\;$ region of BD +23 3130.  The dots are the observed data points 
 (binned 2x) and the lines are the synthetic spectra for the stellar parameters 
 adopted by this paper for [O/Fe] = +0.0, +0.5, +1.0 and +1.5. 

\noindent {\bf Figure 4b}.  Same as Figure 4a, except for BD +37 1458. 

\noindent {\bf Figure 5a}.  Same as Figure 4a, except using the parameters of 
 I98. 

\noindent {\bf Figure 5b}.  Same as Figure 4b, except using the parameters of 
 I98. 

\noindent {\bf Figure 5c}.  Same as Figure 4b, except using the parameters of  
 B99 (King 1993 scale).

\noindent {\bf Figure 5d}.  Same as Figure 4b, except using the parameters of
 B99 (Carney 1983 scale).


\plotone{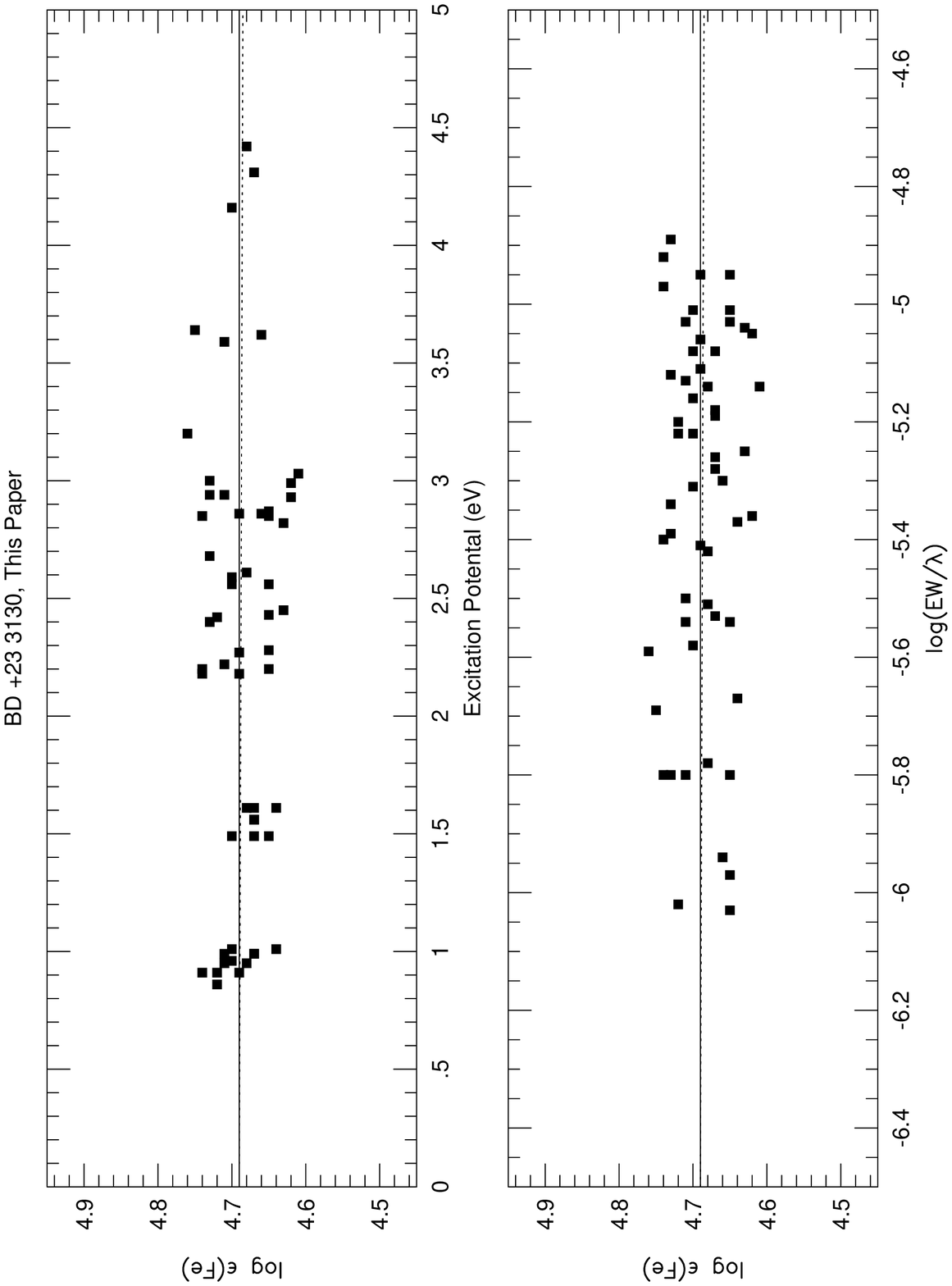}

\plotone{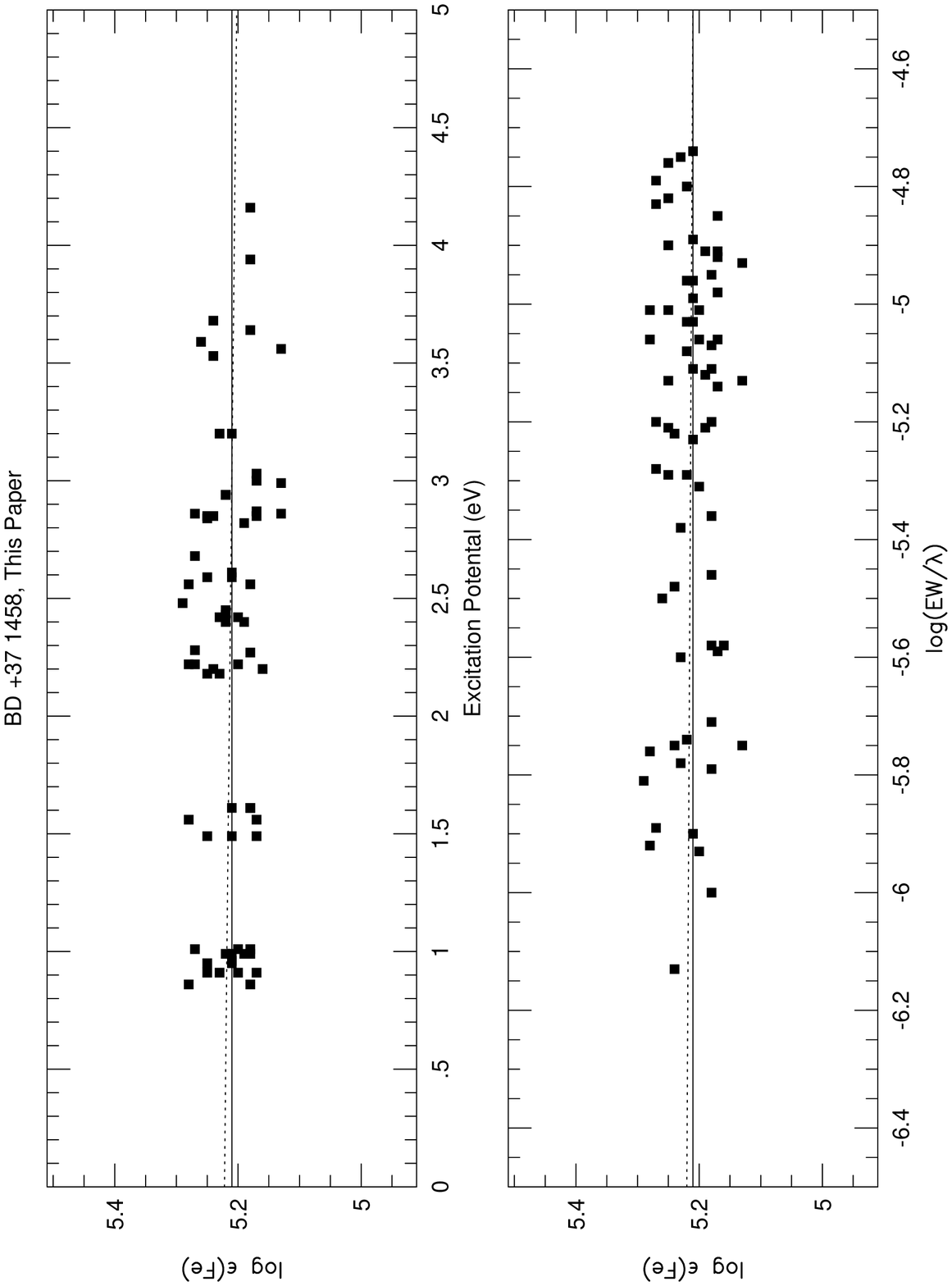}

\plotone{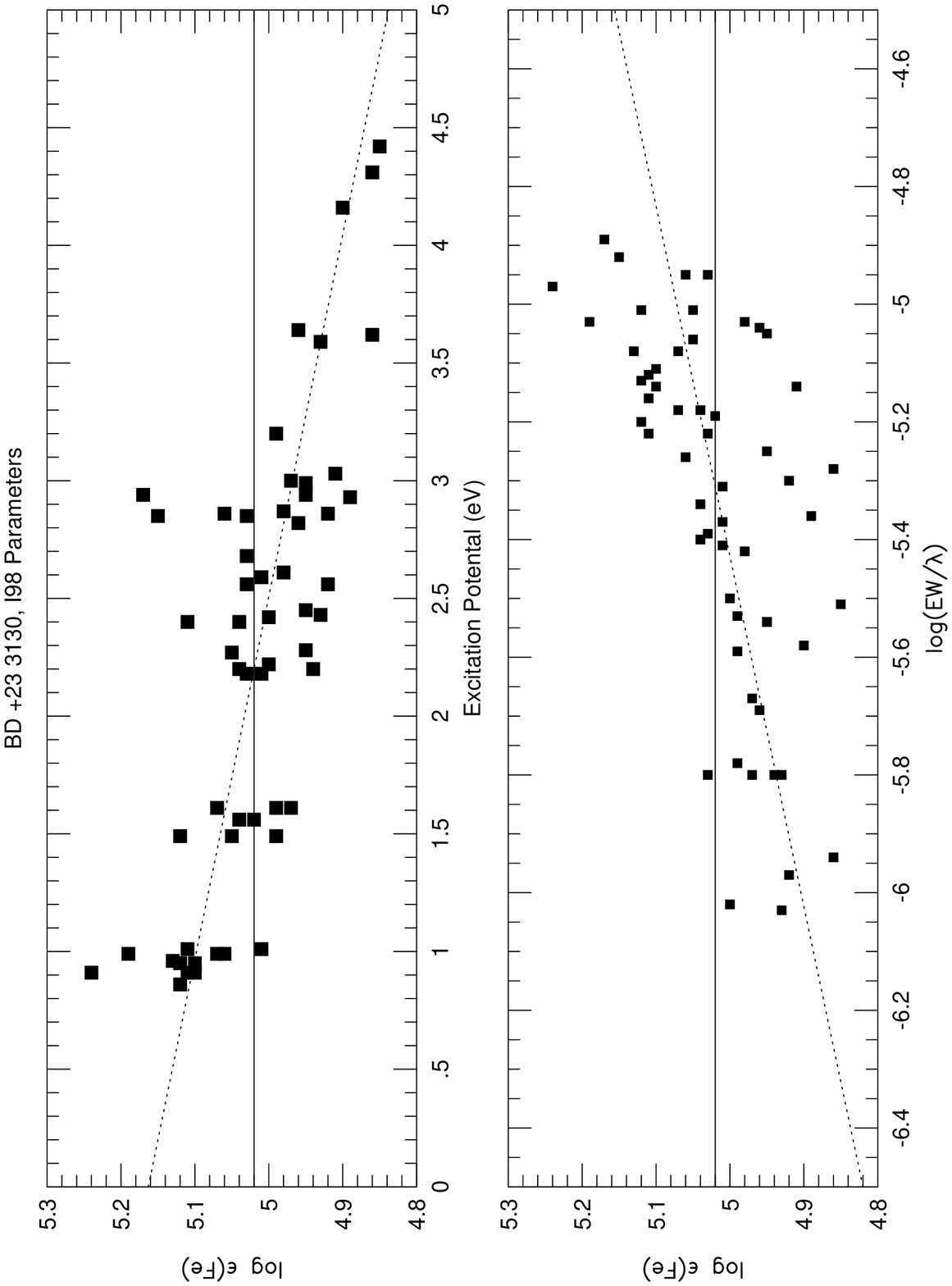}

\plotone{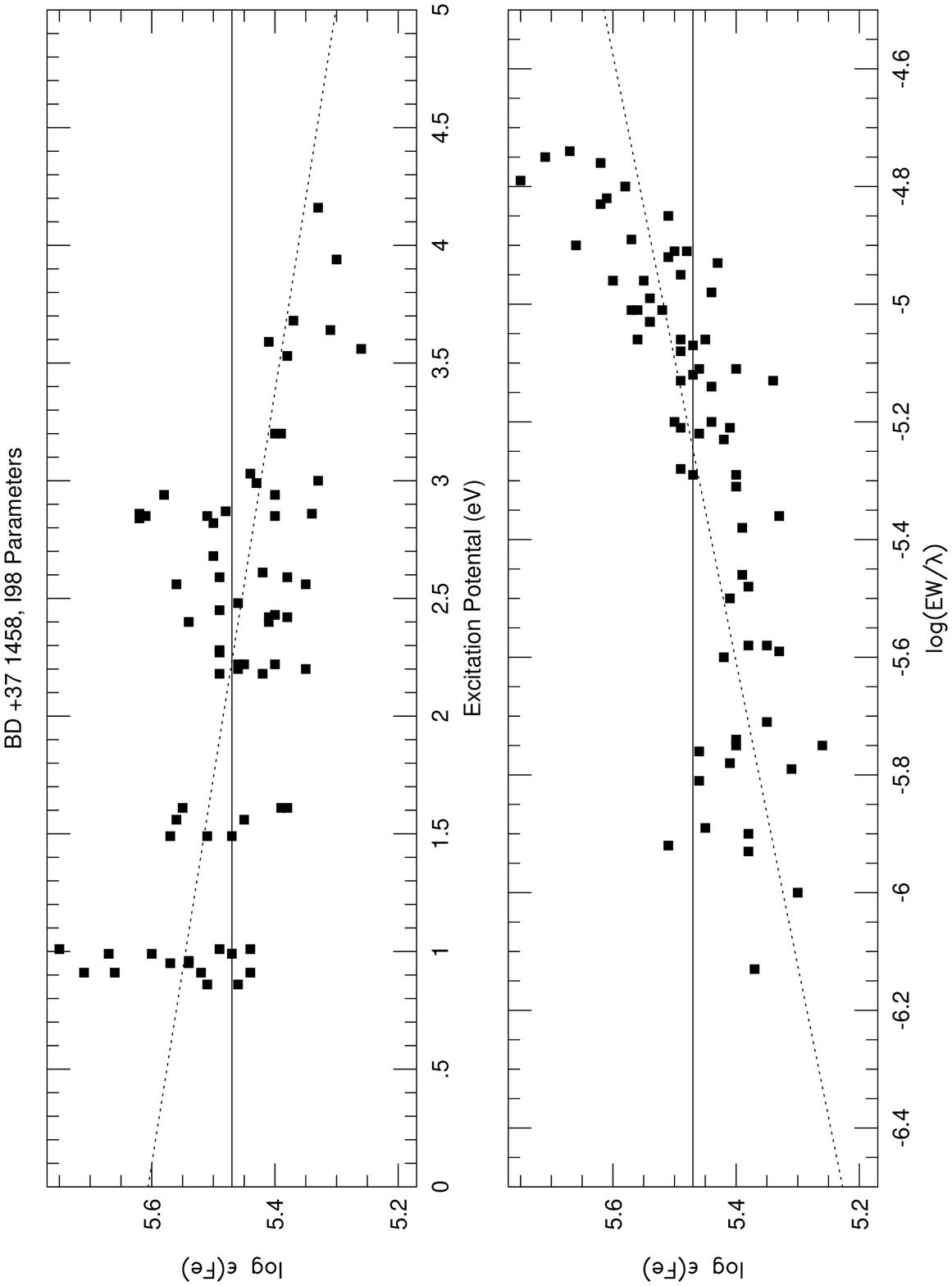}

\plotone{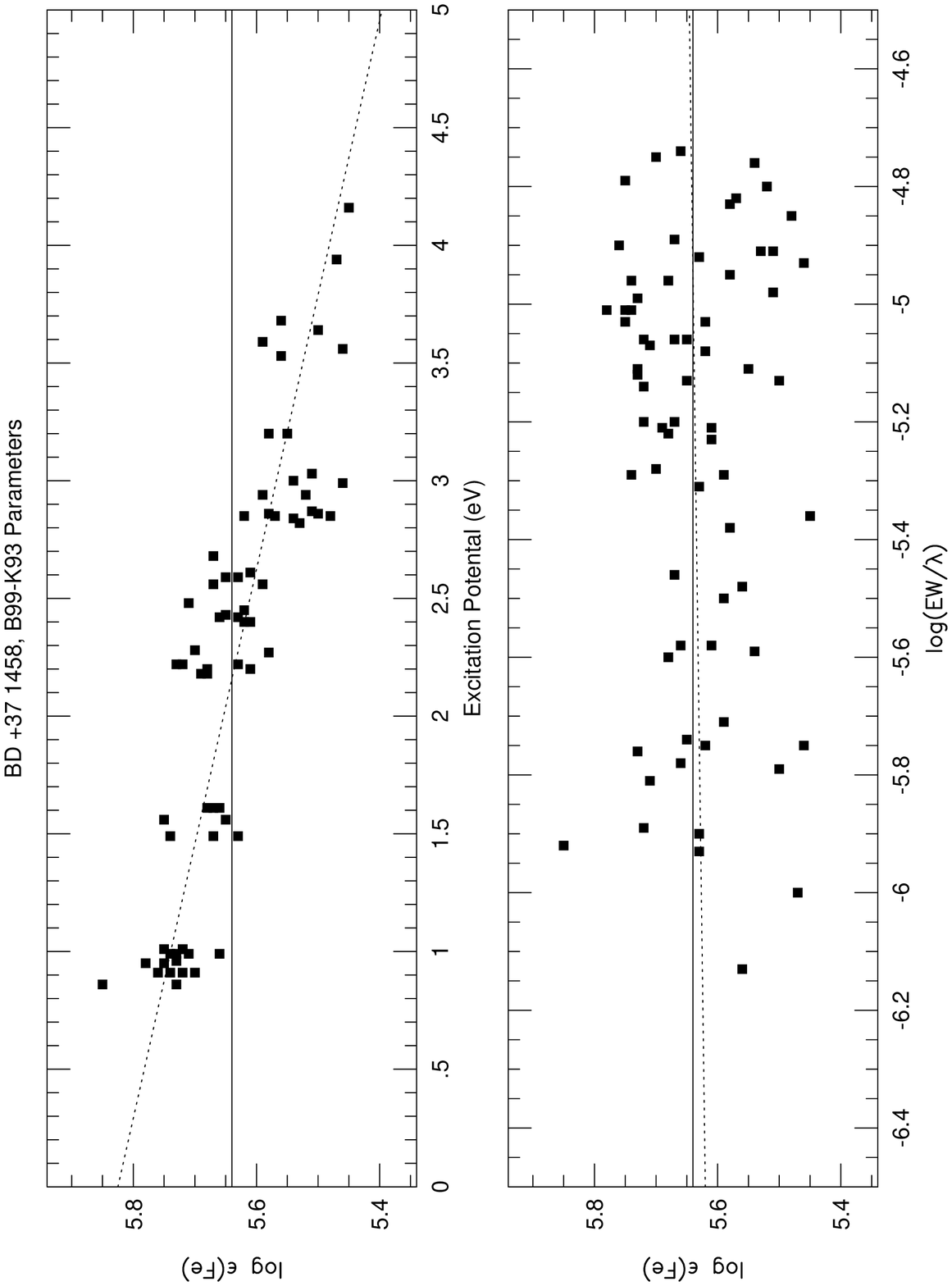}

\plotone{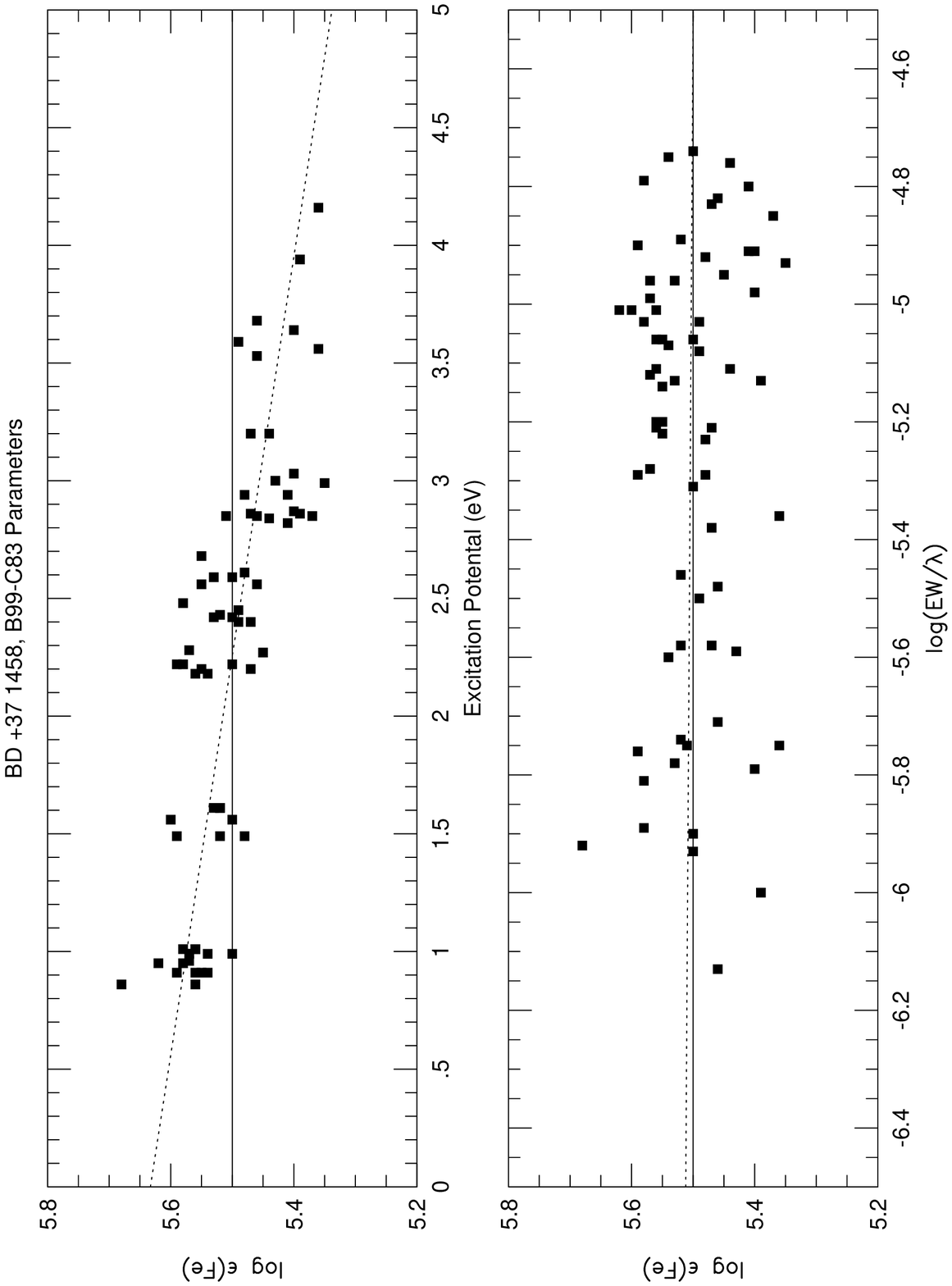}

\plotone{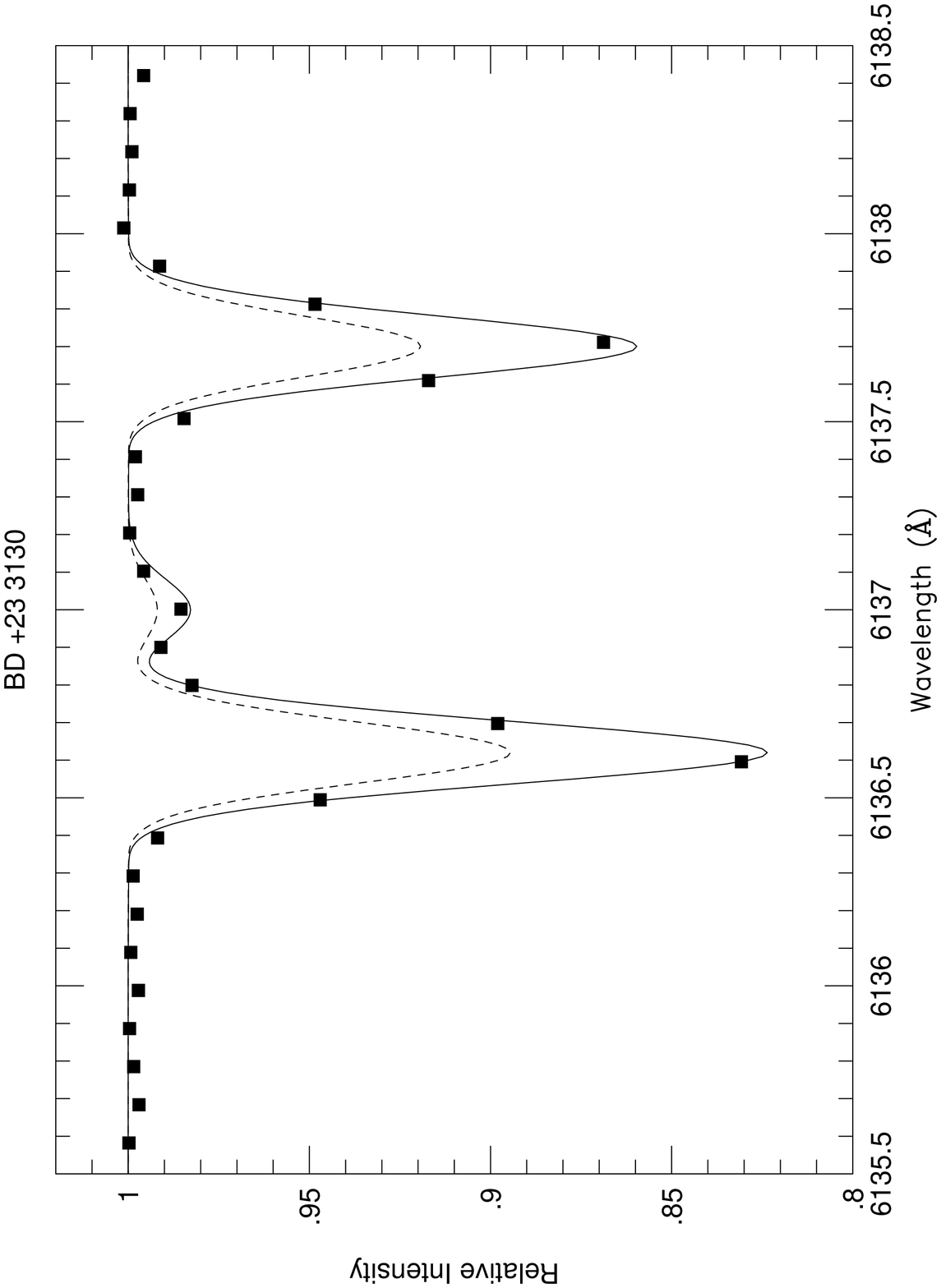}

\plotone{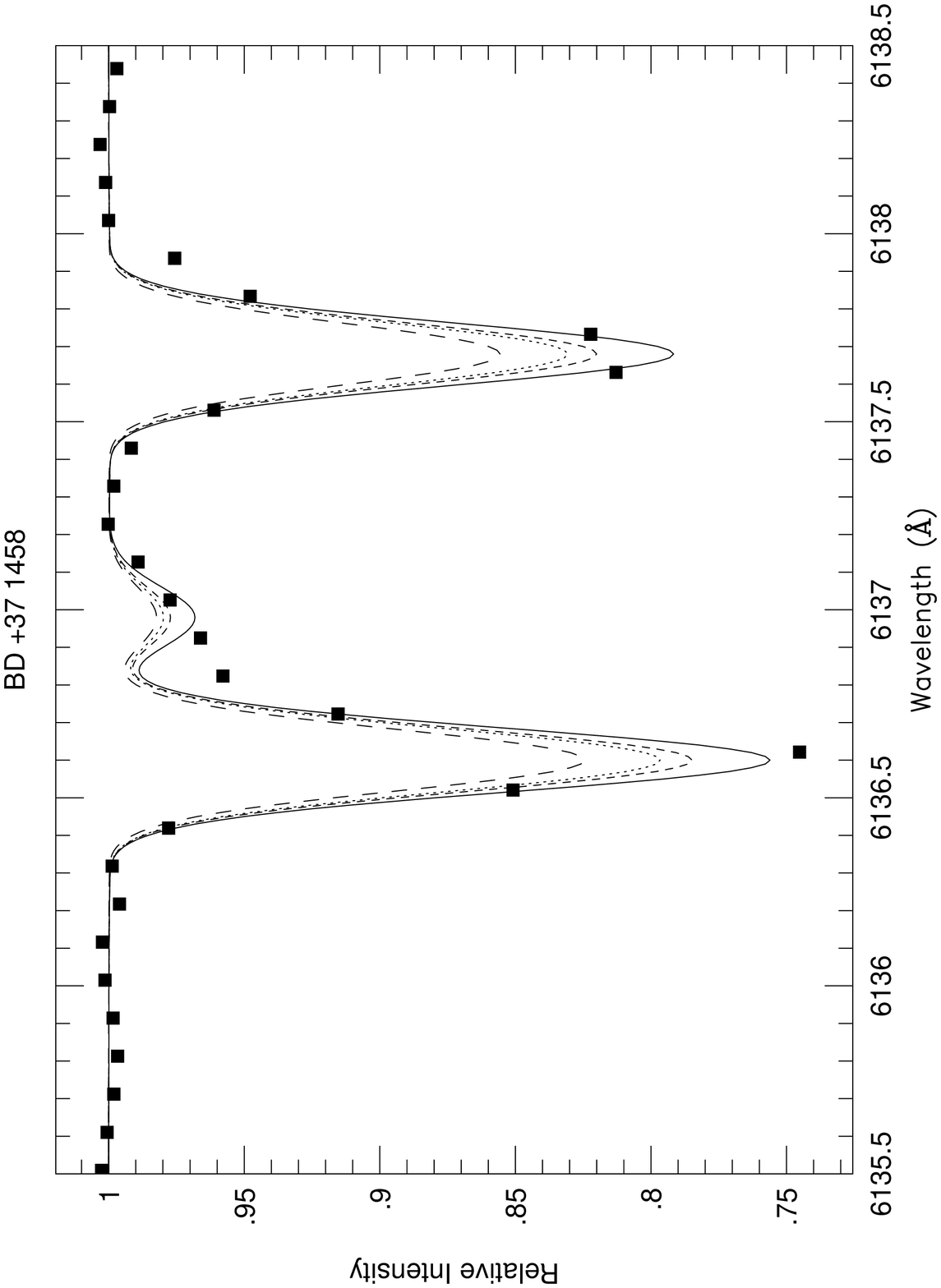}

\plotone{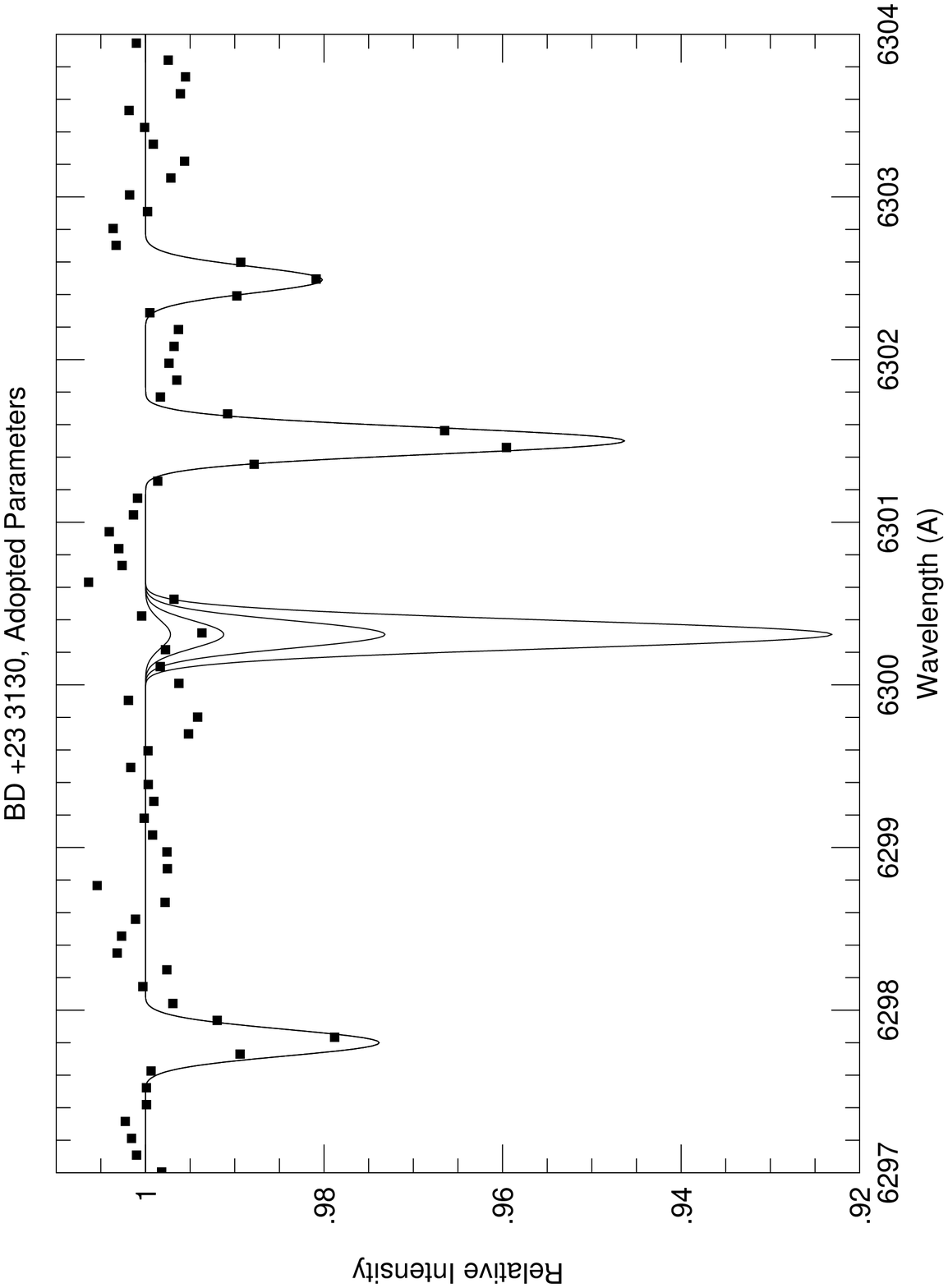}

\plotone{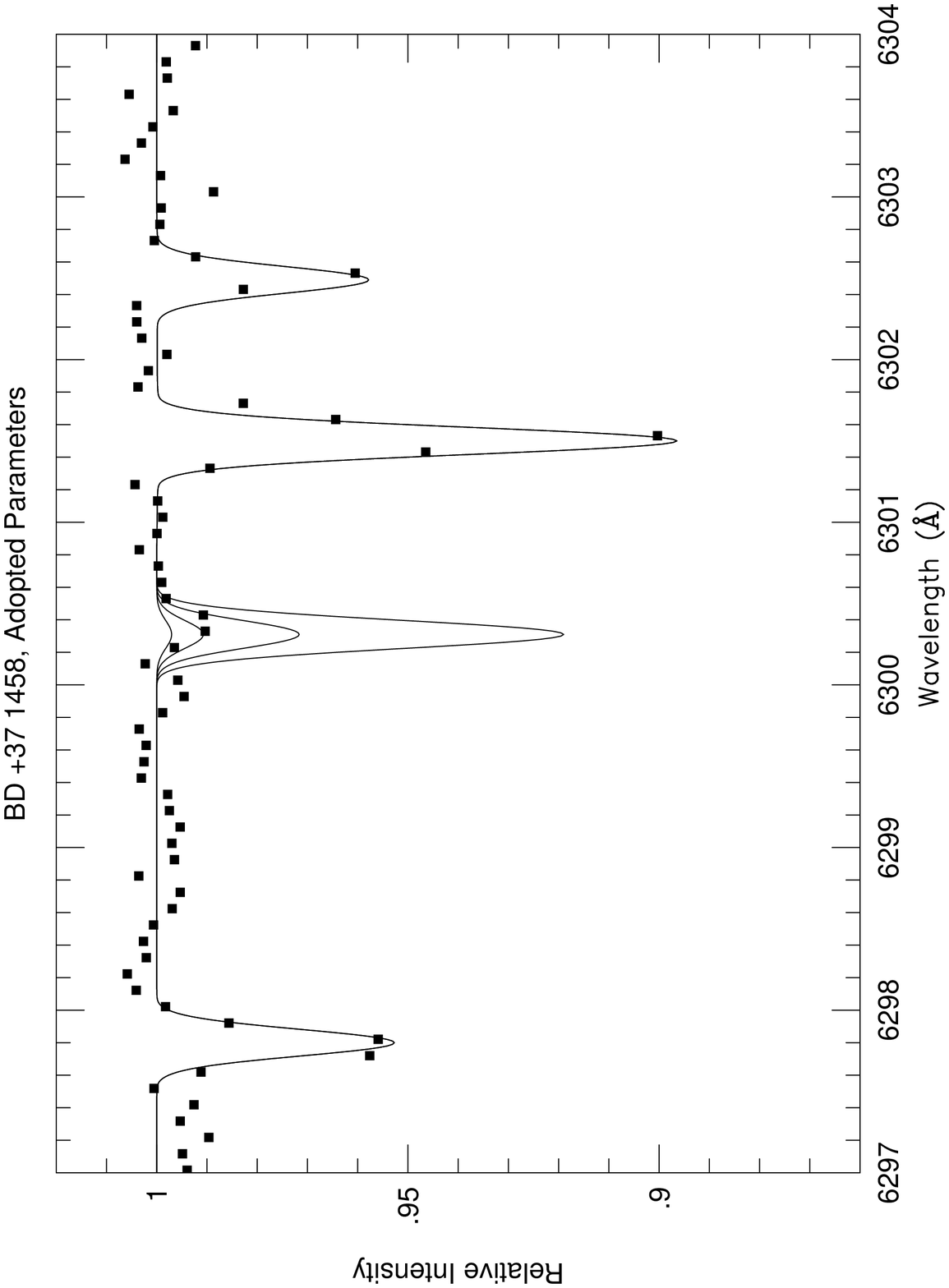}

\plotone{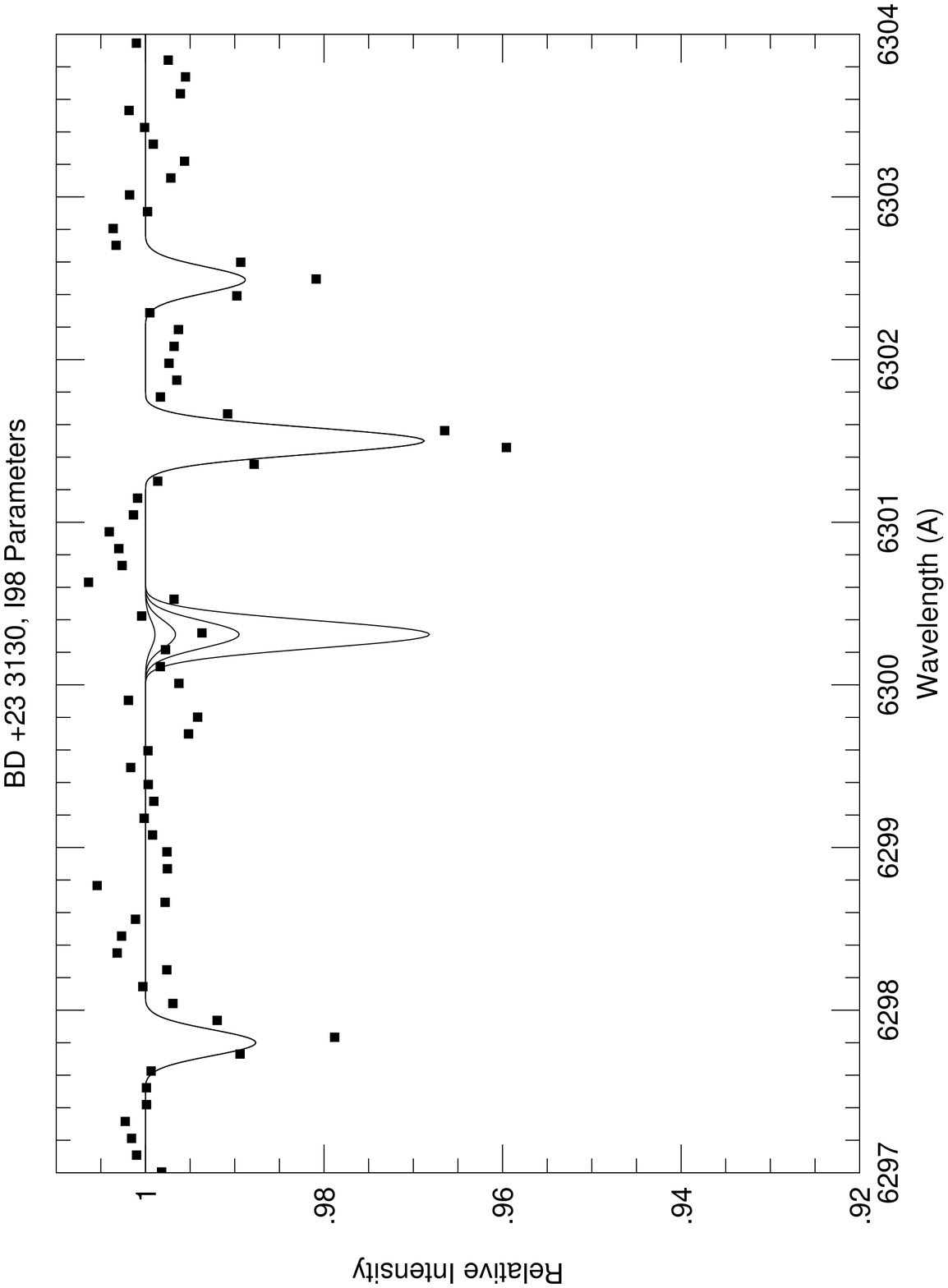}

\plotone{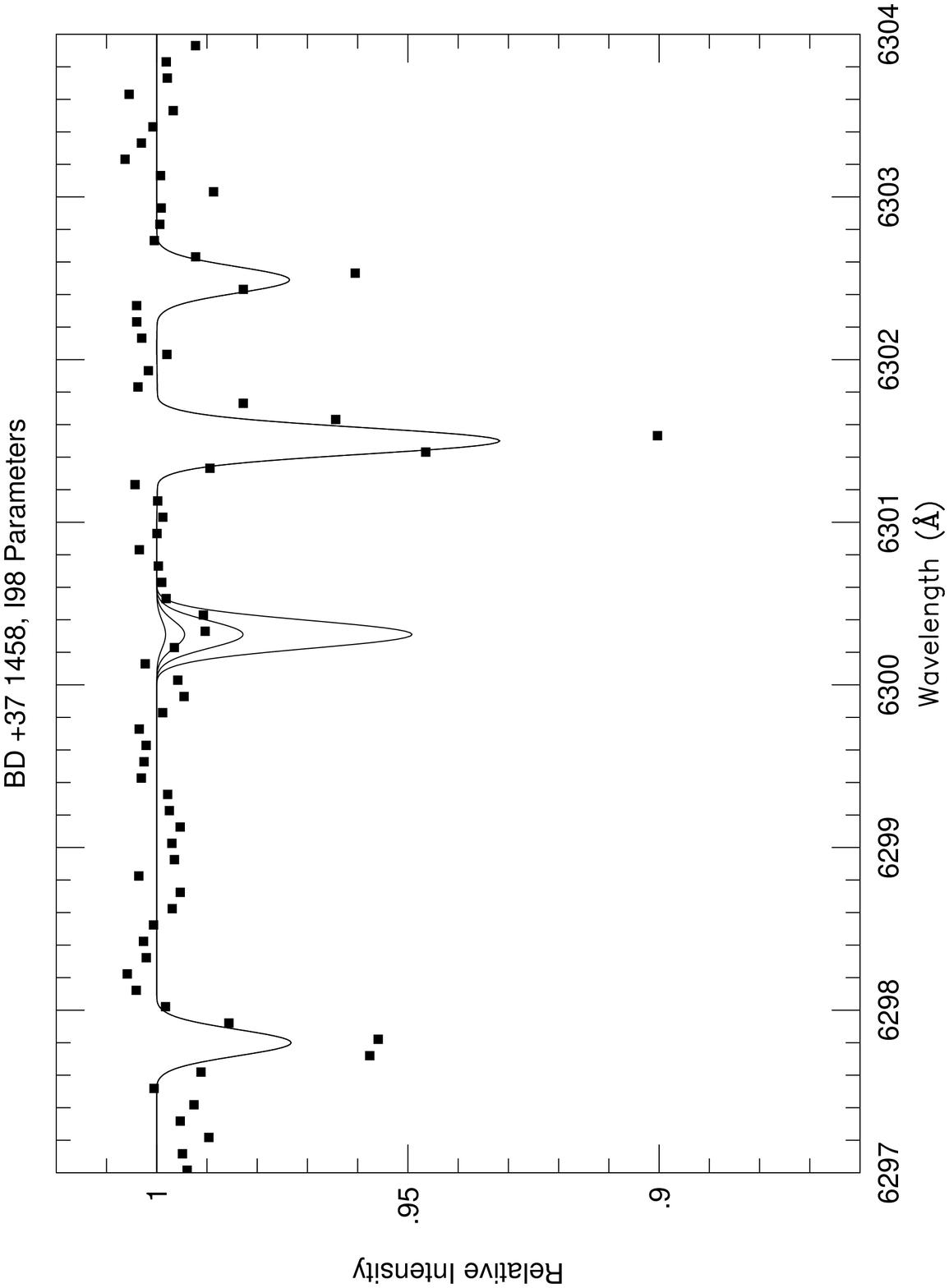}

\plotone{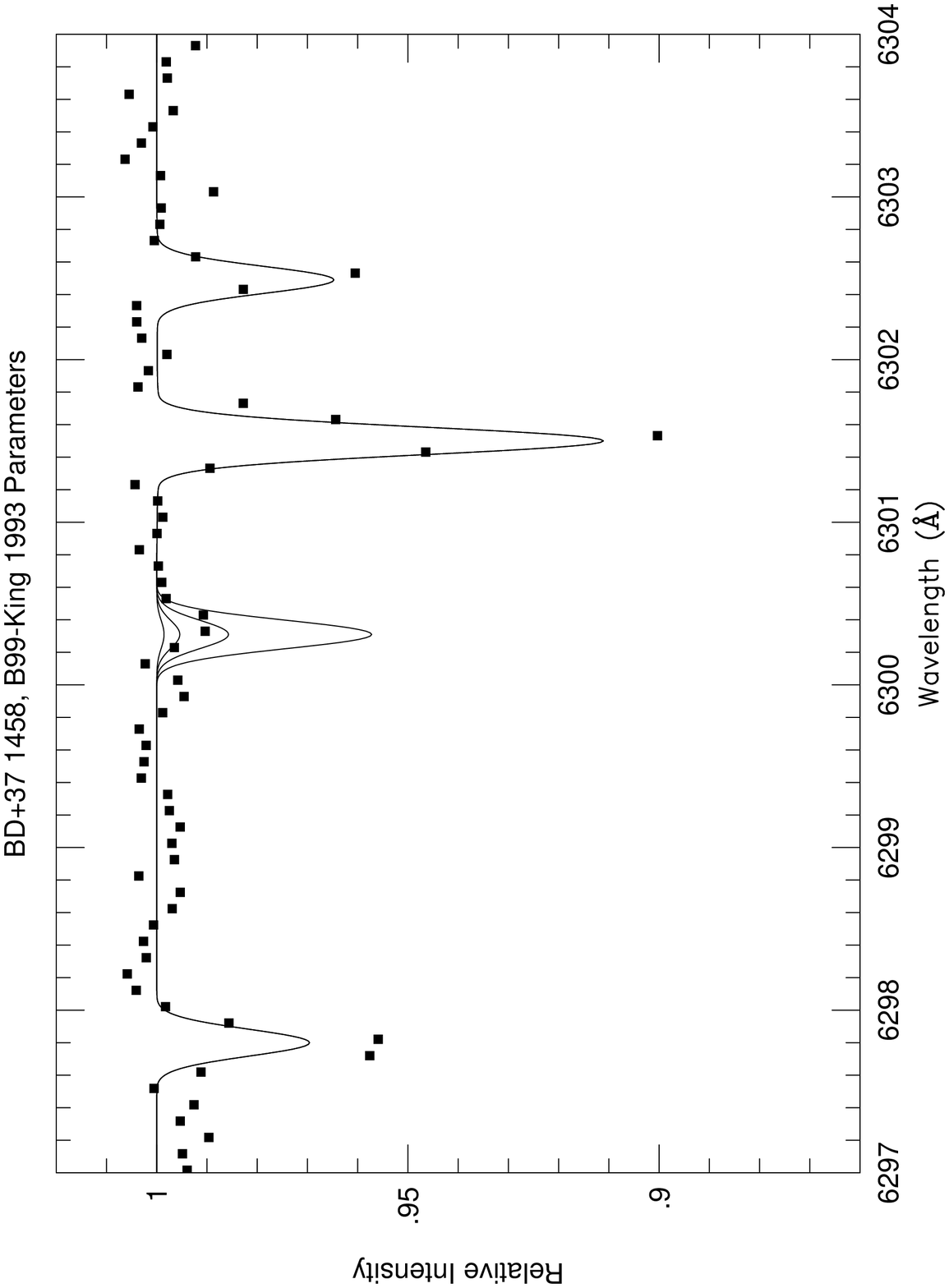}

\plotone{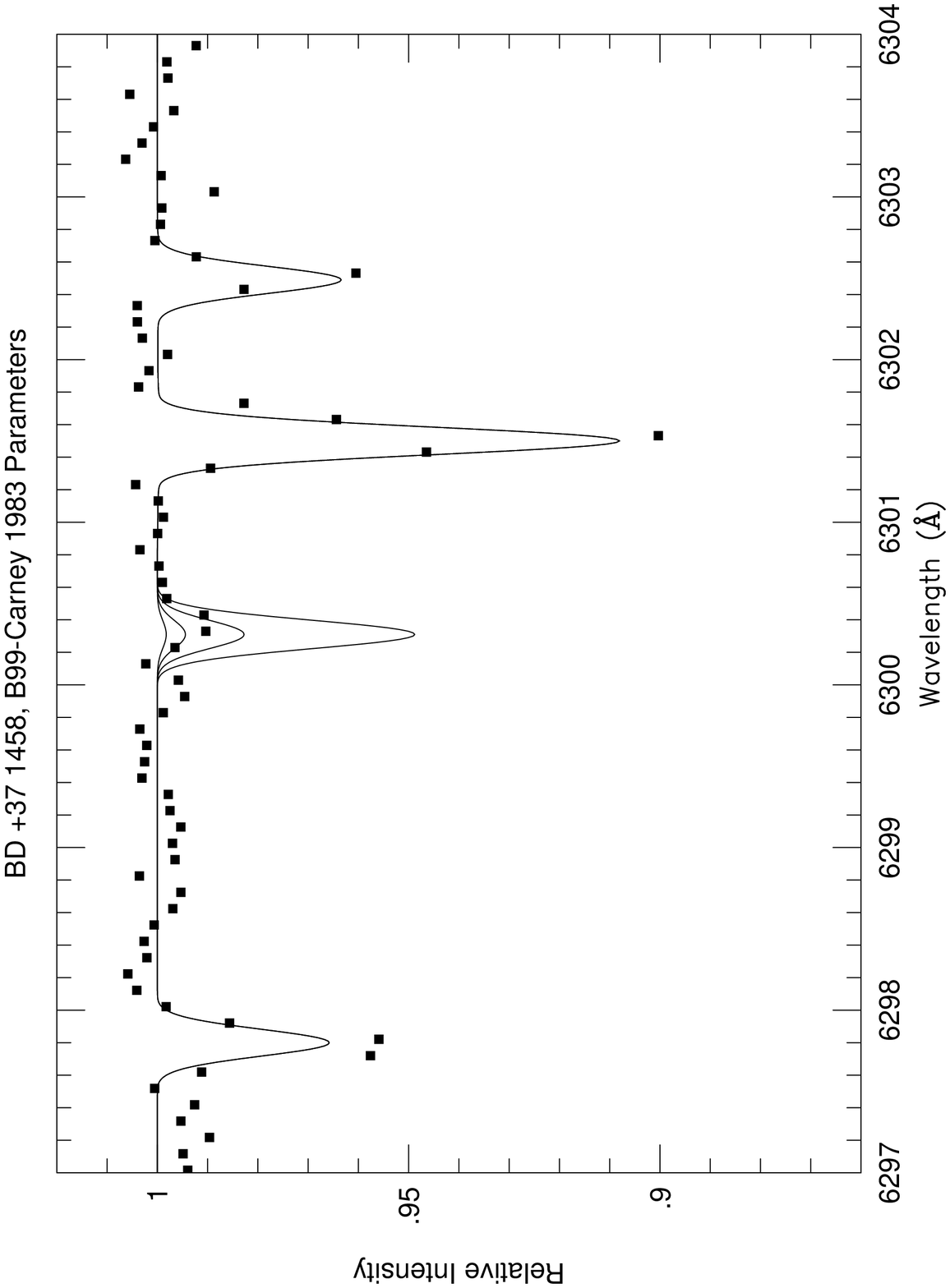}
 



\begin{thebibliography}{}

\bibitem[Abia \& Rebolo 1989]{ar89} Abia, C. \& Rebolo, R. 1989, \apj, 347, 186.

\bibitem[Anders \& Grevesse (1989)]{ag89} Anders, E. \& Grevesse, N. 1989, \gca, 53, 197.

\bibitem[Arnett 1995]{ar95} Arnett, D. 1995, \araa, 33, 115.

\bibitem[Balachandran \& Carney (1996)]{bc96} Balachandran, S. \& Carney, B. W. 1996, \aj, 111, 946.

\bibitem[Bessell et al. 1984]{bhc84} Bessell, M. S., Hughes, S. \& Cottrell, P. L. 1984, Pub. Astr. Soc. Australia,
         5, 547.

\bibitem[Bessell et al. (1991)]{bsr91} Bessell, M. S., Sutherland, R. S. \& Ruan, K. 1991, \apjl, 383, 71.

\bibitem[Beveridge \& Sneden 1994]{bs94} Beveridge, R. C. \& Sneden, C. 1994, \aj, 108, 285.

\bibitem[Biemont et al. (1991)]{b91} Biemont, E., Baudoux, M., Kurucz, R. L., Ansbacher, W. \& Pinnington, E. H. 1991, \aap, 249, 539.

\bibitem[Blackwell et al. 1979a]{b79a} Blackwell, D. E., Ibbetson, P. A., Petford, A. D. \& Shallis, M. J.  1979a, \mnras, 186, 633.

\bibitem[Blackwell et al. 1979b]{b79b} Blackwell, D. E., Petford, A. D. \& Shallis, M. J.  1979b, \mnras, 186, 657.

\bibitem[Blackwell et al. 1980]{b80} Blackwell, D. E., Petford, A. D., Shallis, M. J. \& Simmons, G. J. 1980, \mnras, 191, 445.

\bibitem[Blackwell et al. 1982a]{b82a} Blackwell, D. E., Shallis, M. J. \& Simmons, G. J. 1982a, \mnras, 199, 33.

\bibitem[Blackwell et al. 1982b]{b82b} Blackwell, D. E., Petford, A. D., Shallis, M. J. \& Simmons, G. J. 1982b, \mnras, 199, 43.

\bibitem[Blackwell et al. 1982c]{b82c} Blackwell, D. E., Petford, A. D. \& Simmons, G. J. 1982c, \mnras, 201, 593.

\bibitem[Blackwell et al. 1986]{bl86} Blackwell, D. E., Booth, A. J., Haddock, D. J., Petford, A. D. \& Leggett, S. K. 1986, \mnras, 220, 549.

\bibitem[Boesgaard et al 1999]{b99} Boesgaard, A., King, J. R., Deliyannis, C. P. \& Vogt, S. S. 1999, \aj, 117, 492.

\bibitem[Briley et al. 1994]{b94} Briley, M. M., Bell, R. A., Hesser, J. E. \& Smith, G. H. 1994, Can. J. Phys., 72, 772.
         
\bibitem[Carbon et al. (1982)]{c82} Carbon, D. F., Langer, G. E., Butler, D., Kraft, R. P., Trefzger, C. F., Suntzeff, N. B., Kemper, E. \& Romanishin, W. 1982, \apjs, 49, 207.

\bibitem[Carney (1983)]{c83} Carney, B. W. 1983 \aj, 83 610.

\bibitem[Carney et al. 1994]{clla} Carney, B. W., Latham, D. W., Laird, J. B. \& Aguilar, L. A. 1994, \aj, 107, 2240.

\bibitem[Cavallo et al. 1997]{c97} Cavallo, R. M., Pilachowski, C. A. \& Rebolo, R. 1997, \pasp, 109, 226.

\bibitem[Chaboyer et al. 1998]{ch98} Chaboyer, B., Demarque, P., Kernan, P. J.
 \& Krauss, L. M. 1998, \apj, 494, 96.

\bibitem[Djorgovski 1993]{d93} Djorgovski, S. 1993, in ASP Conf. Ser., 50, 373.

\bibitem[Fuhr et al. 1988]{f88} Fuhr, J. R., Martin, G. A. \& Weise, W. L. 1988, J. Phys. Chem. Ref. Data, 17, Suppl. No. 4, 1.

\bibitem[Fuhrmann et al. 1995]{f95} Fuhrmann, K., Axer, M. \& Gehren, T. 1995, \aap, 301, 492.

\bibitem[Hipparchos Survey (1997)]{hip} ESA 1997.  The Hipparchos \& Tycho Catalogues (ESA SP-1200) (Noordwijk: ESA).

\bibitem[Garz 1973]{g73} Garz, T. 1973, \aap, 26, 471.

\bibitem[Gonzalez \& Wallerstein 1998]{gw98} Gonzalez, G. \& Wallerstein, G. 1998, \aj (in press).

\bibitem[Hanson et al. 1998]{h98} Hanson, R. B., Sneden, C., Kraft, R. P. \& Fulbright, J. P. 1998, \aj, 116, 1286.

\bibitem[Hoffleit 1964]{bsc} Hoffleit, D., 1964, Bright Star Catalog, Yale University Observatory 

\bibitem[Israelian et al. 1998]{i98} Israelian, G., Garcia Lopez, R. J. \& Rebolo, R. 1998, \apj, 507, 805.

\bibitem[King (1993)]{k93} King, J. R. 1993, \aj, 106, 1206.

\bibitem[Kisselman \& Nordlund 1995]{kn95} Kisselman, D. \& Nordlund, A. 1995, \aap, 302, 578.

\bibitem[Kraft 1994]{k94} Kraft, R. P. 1994, \pasp, 106, 553.

\bibitem[Kraft et al. 1997]{k97} Kraft, R. P., Sneden, C., Smith, G. H., Shetrone, M. D., Langer, G. E. \& Pilachowski, C. A. 1997, \aj, 113, 279.

\bibitem[Kroll \& Kock (1987)]{kk87} Kroll, S. \& Kock, M. 1987, \aaps, 67, 225.

\bibitem[Kurucz (1992)]{k92} Kurucz, R. L. 1992, private communication. 

\bibitem[Lambert 1978]{l78} Lambert, D. L. 1978, \mnras, 183, 249.

\bibitem[Lambert \& Warner 1968]{lw68} Lambert, D. L. \& Warner, B. 1968, \mnras, 138, 181.

\bibitem[Martin et al. 1988]{m88} Martin, G. A., Fuhr, J. R. \& Wiese, W. L. 1988, J. Phys. Chem. Ref. Data, 17, Suppl. No. 3, 1.

\bibitem[McWilliam 1997]{m97} McWilliam, A. 1997, \araa, 35, 503.

\bibitem[McWilliam et al. (1995)]{m95} McWilliam, A., Preston, G. W., Sneden, C. \& Searle, L. 1995, \aj, 109, 2757.

\bibitem[Nissen et al. (1994)]{n94} Nissen, P. E., Gustafsson, B., Edvardsson, B. \& Gilmore, G. 1994, \aap, 285, 440.

\bibitem[O'Brian et al. (1991)]{o91} O'Brian, T. R., Wickliffe, H. E., Lawler, J. E., Whaling, W. \& Brault, J. W. 1991, J. Opt. Soc. Am. B, 8, 1185.

\bibitem[Rood \& Crocker 1985]{rc85} Rood, R. T. \& Crocker, D. A. 1985, in ESO Workshop on Production and Distribution of C, N, O Elements, ed. I. J. Danziger, F. Matteucci, \& K. Kjar (ESO, Garching), p. 61.

\bibitem[Schlegel et al. 1998]{sfd98} Schlegel, D. J., Finkbeiner, D. P. \& Davis, M. 1998, \apj, 500, 525.

\bibitem[Schuster \& Nissen (1989)]{sn89} Schuster, W. J. \& Nissen, P. E. 1989, \aap, 222, 69.

\bibitem[Shetrone 1996]{matt96} Shetrone, M. D. 1996, \aj, 112, 1517.

\bibitem[Smith \& Raggett 1981]{sr81} Smith, G. \& Raggett, D. S. J. 1981, J. Phys. B. 14, 4015.

\bibitem[Sneden 1973]{s73} Sneden, C. 1973, \apj, 184, 839.

\bibitem[Sneden et al. 1991]{s91} Sneden, C., Kraft, R. P., Prosser, C. F. \& Langer, G. E. 1991, \aj, 102, 2001.

\bibitem[Sneden et al. (1994)]{s94} Sneden, C., Kraft, R. P., Langer, G. E., Prosser, C. F. \& Shetrone, M. D. 1994, \aj, 107, 1773.
         
\bibitem[Sneden et al. 1997]{s97} Sneden, C., Kraft, R. P., Shetrone, M. D., Smith, G. H., Langer, G. E. \& Prosser, C. F. 1997, \aj, 114, 1964.
         
\bibitem[Stetson \& Harris 1988]{sh88} Stetson, P. B. \& Harris, W. E. 1988, \aj, 96, 909.

\bibitem[Straniero \& Chieffi 1991]{sc91} Straniero, O. \& Chieffi, A. 1991, \apjs, 76, 525.

\bibitem[Suntzeff 1993]{sun93} Suntzeff, N. B. 1993, in ASP Conf. Ser., 48, 167.

\bibitem[Takeda 1995]{t95} Takeda, Y. 1995, \pasj, 47, 463.

\bibitem[Thevenin 1989]{t89} Thevenin, F.  1989, \aaps, 77, 137.

\bibitem[Tomkin et al. 1992]{t92} Tomkin, J., Lemke, M., Lambert, D. L. \& Sneden, C. 1992, \aj, 104, 1568.

\bibitem[VandenBerg 1985]{v85} VandenBerg, D. A. 1985, in ESO Workshop on Production and Distribution of
         C, N, O Elements, ed. I. J. Danziger, F. Matteucci \& K. Kjar
         (ESO, Garching), p. 73.

\bibitem[VandenBerg (1992)]{v92} VandenBerg, D. A. 1992, \apj, 391, 685.

\bibitem[Vogt 1987]{v87} Vogt, S. S. 1987, \pasp, 99, 1214.

\bibitem[Weise et al. 1969]{w69} Weise, W. L., Smith, M. W. \& Miles, B. M. 1969, NBS Ref. Data. Ser.

\bibitem[Weise \& Martin 1980]{wm80} Weise, W. L. \& Martin, G. A. 1980, NSRDS-NBS 68 (U.S. Gov. Printing Office, Washington).

\bibitem[Zinn \& West 1984]{zw84} Zinn, R. \& West, M. J. 1984, \apjs, 55, 45.

\end{thebibliography}
\end{document}